\documentclass[usenatbib,usegraphicx]{mn2e}
\usepackage{amsfonts}
\usepackage{amsmath}
\usepackage{amssymb}
\usepackage{comment}
\usepackage[usenames,dvipsnames]{color}

\title[AGN-galaxy co-evolution at low redshift]
    {The triggering of local AGN and their role in regulating star formation}
\author[Sugata Kaviraj]
{Sugata Kaviraj\thanks{s.kaviraj@herts.ac.uk}$^{1}$, Stanislav
S. Shabala$^{2}$, Adam T. Deller$^{3}$ and Enno Middelberg$^{4}$\\
$^{1}$Centre for Astrophysics Research, University of
Hertfordshire, College Lane, Hatfield, Herts, AL10 9AB, UK\\
$^{2}$School of Mathematics and Physics, University of Tasmania, Private Bag 37, Hobart, TAS 7001, Australia\\
$^{3}$The Netherlands Institute for Radio Astronomy (ASTRON),
Dwingeloo, The Netherlands\\
$^{4}$Astronomisches Institut der Ruhr-Universit\"{a}t Bochum,
Universit\"{a}tsstra\ss e 150, D-44801 Bochum, Germany}

\begin{document}

\maketitle

\def \aj {AJ}
\def \mnras {MNRAS}
\def \pasp {PASP}
\def \apj {ApJ}
\def \apjs {ApJS}
\def \apjl {ApJL}
\def \aap {A\&A}
\def \nat {Nature}
\def \araa {ARAA}
\def \iaucirc {IAUC}
\def \aaps {A\&A Suppl.}
\def \qjras {QJRAS}
\def \na {New Astronomy}
\def \aapr {A\&ARv}
\def\lesssim{\mathrel{\hbox{\rlap{\hbox{\lower4pt\hbox{$\sim$}}}\hbox{$<$}}}}
\def\gtrsim{\mathrel{\hbox{\rlap{\hbox{\lower4pt\hbox{$\sim$}}}\hbox{$>$}}}}


\begin{abstract}
We explore the processes that trigger local AGN and the role of
these AGN in regulating star formation, using $\sim$350 nearby
galaxies observed by the mJy Imaging VLBA Exploration at 20cm
(mJIVE) survey. The $\gtrsim 10^7$ K brightness temperature
required for an mJIVE detection cannot be achieved via star
formation alone, allowing us to unambiguously detect nearby radio
AGN and study their role in galaxy evolution. Radio AGN are an
order of magnitude more common in early-type galaxies (ETGs) than
in their late-type counterparts. The VLBI-detected ETGs in this
study have a similar stellar mass distribution to their undetected
counterparts, are typically not the central galaxies of clusters
and exhibit merger fractions that are significantly higher than in
the average ETG. This suggests that these radio AGN (which have
VLBI luminosities $>10^{22}$ W Hz$^{-1}$) are primarily fuelled by
mergers, and not by internal stellar mass loss or cooling flows.
Our radio AGN are a factor of $\sim$3 times more likely to reside
in the UV-optical red sequence than the average ETG. Furthermore,
typical AGN lifetimes (a few $10^7$ yr) are much shorter than the
transit times from blue cloud to red sequence ($\sim$1.5 Gyr).
This indicates that the AGN are not triggered promptly and appear
several dynamical timescales into the associated star formation
episode, implying that they typically couple only to residual gas,
at a point where star formation has already declined
significantly. While evidence for AGN feedback is strong in
systems where the black hole is fed by the cooling of hot gas, AGN
triggered by mergers appear not to strongly regulate the
associated star formation. The inability of the AGN to rapidly
quench merger-driven star formation is likely to make merging the
dominant mode of star formation in nearby ETGs, in line with the
growing evidence for minor mergers being the primary driver of
stellar mass growth in these systems at low redshift.
\end{abstract}




\section{Introduction}
Understanding the assembly of massive galaxies, which dominate the
stellar mass density in today's Universe (e.g. Kaviraj 2014a), is
a fundamental topic in observational cosmology. While simple
physics, implemented within the $\Lambda$CDM paradigm,
successfully reproduces many observed properties of today's
galaxies
\citep[e.g.][]{Cole2000,Hatton2003,Springel2005a,Croton2006,Somerville2012},
it has been recognized for some time that basic characteristics,
such as the distribution of luminosities and colours, cannot be
reproduced without invoking energetic feedback that quenches star
formation at both ends of the galaxy mass function
\citep{Benson2003,Shabala2009}.

In low-mass galaxies ($M_* \lesssim 10^{10} M_{\odot}$, see
Kaviraj et al. 2007a), typical energies imparted by supernovae
ejecta are sufficient to remove gas from the shallow potential
wells and quench star formation. In the high-mass regime, however,
the gravitational potential wells are too deep for supernovae to
be effective, and a more energetic source of feedback is required
\citep[e.g][]{Silk1998}. An attractive source of this feedback is
the central black hole, because the potential energy released by
the growth of the black hole is several orders of magnitude larger
than the binding energy of the gas reservoir, even in the most
massive galaxies \citep[e.g.][]{Fabian2012}.

Our theoretical picture of galaxy evolution postulates a key role
for AGN in regulating star formation in massive galaxies across
cosmic time. At high-redshift, where most of today's stellar mass
was assembled in intense star formation episodes, black holes were
likely to be accreting close to the Eddington limit. Quasar-driven
winds from such accretion episodes are postulated to remove gas
reservoirs and truncate star formation after a few dynamical
timescales
\citep[e.g.][]{Silk1998,Haehnelt1998,Fabian1999,King2003,Springel2005,Fabian2012}.
Notwithstanding this early removal of gas, a massive galaxy will
retain the ability to accrete fresh gas, which may continue to
fuel star formation. There are several potential sources of this
late-stage accretion, such as mergers, stellar mass loss (which
feeds the internal hot gas reservoir) and cooling flows on to the
central galaxies of clusters. The quiescence of massive galaxies
must therefore be \emph{maintained}, the current theoretical
consensus favouring feedback from AGN to perform this regulation
over the latter half of cosmic time.

Observational work lends support to this theoretical picture. The
peak of the cosmic star formation and black hole accretion rate
densities coincide at $z\sim2$ \citep[e.g.][]{Madau2014}, while in
the local Universe galaxy (bulge) mass (M$_{\textnormal{GAL}}$)
and black hole mass (M$_{\textnormal{BH}}$) show a strong
correlation \citep[see
e.g.][]{Magorrian1998,Gebhardt2000,Ferrarese2000,Haring2004,Gultekin2009,McConnell2011},
with M$_{\textnormal{GAL}}$ $\sim$ M$_{\textnormal{BH}}$ $\times$
1000. Note that, while this correlation suggests co-evolution
between the two systems
\citep[e.g.][]{Silk1998,Granato2004,Springel2005,Croton2006}, it
is possible that this scaling relation could be a natural
consequence of hierarchical growth via galaxy merging, from
initially uncorrelated distributions of black hole and stellar
masses \citep[e.g.][]{Peng2007,Jahnke2011}.

In some individual galaxies, AGN-driven molecular outflows have
been reported, both in the nearby Universe and at high redshift.
While it is often difficult to attribute molecular outflows
uniquely to the AGN as opposed to the starburst itself, the
energetics of some outflows require the AGN to play a role, since
the outflow rate far exceeds what can be driven by star formation
alone
\citep[e.g.][]{Nesvadba2008,Nesvadba2011,Alexander2010,Rupke2011,Sturm2011,Nyland2013,Morganti2013}.
Perhaps the strongest evidence for AGN feedback comes from the
central galaxies of clusters, which are surrounded by hot gas with
short cooling times \citep{Fabian1994}. Unless the initial mass
function is dominated by low-mass stars
\citep[e.g.][]{Cappellari2012}, as could be possible in the
high-pressure environment of a cluster centre \citep{Fabian1982},
the derived star formation rates in these systems appear to be
several orders of magnitudes lower than the mass deposition rates
expected from cooling. This suggests that the high temperature of
the gas is being maintained, a plausible heat source being the
central black hole
\citep[e.g.][]{Tabor1993,McNamara2007,Cattaneo2009,Fabian2012}.

While individual examples of the interaction of AGN with their
host galaxies are being found, an understanding of the
\emph{global} role of black holes in regulating star formation
requires an analysis of survey-scale samples of AGN. Recent work
using optical emission-line AGN, drawn from the Sloan Digital Sky
Survey \citep[SDSS; ][]{Abazajian2009}, has demonstrated a
coincidence of rising AGN activity and declining star formation
rates, with the peak of the optical AGN activity appearing to lag
behind the peak of the star formation by a few hundred Myrs. This
time lag is observed across the full spectrum of star formation
activity, from the strongly star forming luminous infrared
galaxies \citep{Kaviraj2009a} to the more weakly star forming
early-type galaxies \citep{Schawinski2007,Wild2010,Shabala2012}.
As we explore in our analysis below, a time delay in the onset of
the AGN has important implications for its ability to regulate
star formation, since the gas reservoir may have been
significantly depleted before the AGN has a chance to couple to
it.

\begin{figure}
$\begin{array}{c}
\includegraphics[width=3.3in]{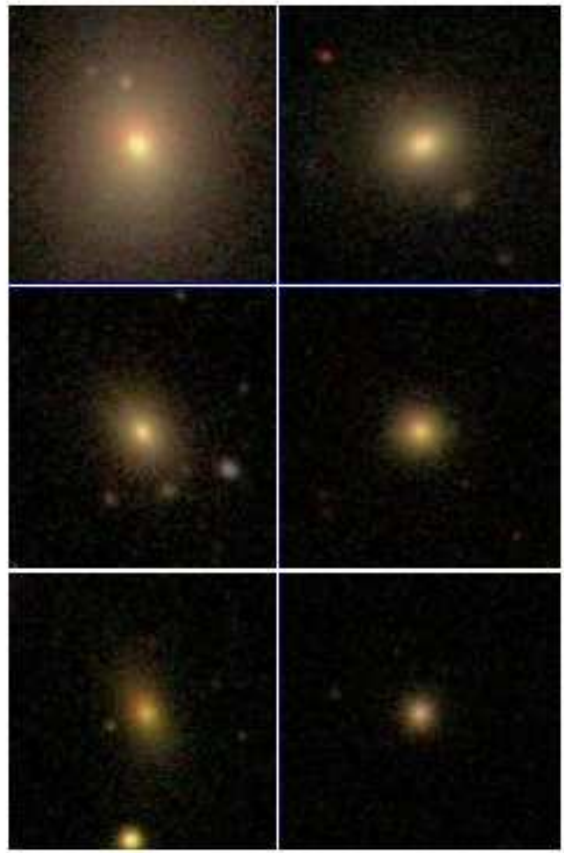}
\end{array}$
\caption{Example SDSS images of early-type mJIVE targets. The
galaxies in the top row are in the redshift range $z<0.1$, those
in the middle row are in the redshift range $0.1<z<0.2$ and those
in the bottom row are in the redshift range $0.2<z<0.3$.}
\label{fig:etg_examples}
\end{figure}

\begin{figure}
\begin{center}
$\begin{array}{c}
\includegraphics[width=3.3in]{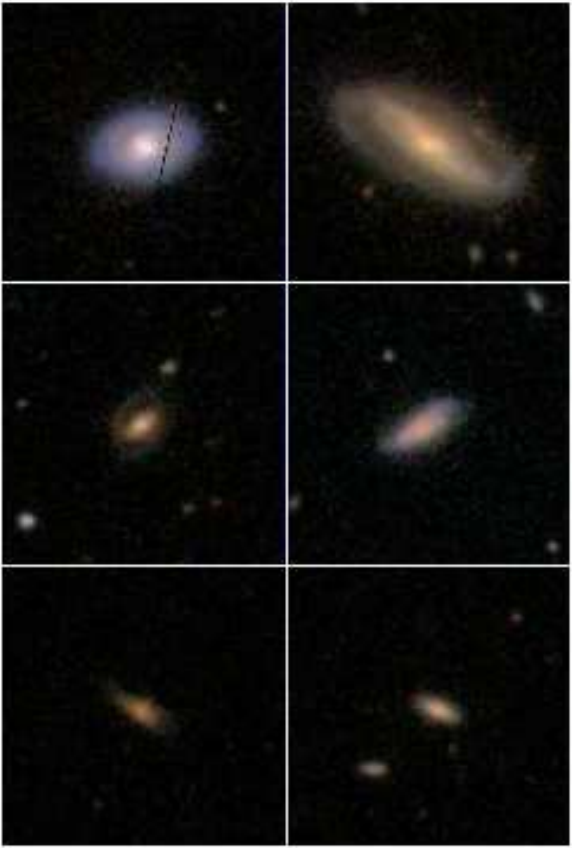}
\end{array}$
\caption{Example SDSS images of late-type mJIVE targets. The
galaxies in the top row are in the redshift range $z<0.1$, those
in the middle row are in the redshift range $0.1<z<0.2$ and those
in the bottom row are in the redshift range $0.2<z<0.3$.}
\label{fig:ltg_examples}
\end{center}
\end{figure}

A drawback of past survey-scale studies is that most datasets are
not able to unambiguously determine the presence of a radio AGN,
which drives the putative maintenance mode feedback. While surveys
like the SDSS can identify optical emission-line AGN, optical and
radio AGN often show little correlation \citep{Best2005}, making
it difficult to establish the presence of the radio mode using
emission lines alone. Furthermore, radio surveys like FIRST and
NVSS typically do not resolve the galaxy core, making it difficult
to ascertain how much of the radio emission is driven by star
formation and how much of it is attributable to a radio AGN.

Unlike the methods discussed above, very long baseline
interferometry (VLBI) in the radio wavelengths can unambiguously
identify AGN, because the high resolution requires brightness
temperatures of $\sim$$10^6$ K for a detection. Such conditions
cannot be achieved via star formation alone and require
non-thermal sources, such as supernova remnants (SNRs), radio
supernovae (SNe) or AGN. However, only in the very local Universe,
and in rather extreme cases like Arp 220 \citep{Lonsdale2006}, can
clusters of luminous SNRs or SNe be sufficiently bright for a VLBI
detection. In the case of this particular study, the galaxies that
will underpin our analysis are early-type systems which have
significantly lower star formation rates (SFRs) than Arp 220,
which has an SFR of several hundred solar masses per year
\citep[e.g.][]{Iwasawa2005,Baan2007}. They also lie much further
away than Arp 220 (see Figure 3), which has a redshift of 0.018.
The compact (parsec-scale) radio emission in these systems is,
therefore, inconsistent with originating from star formation.

The power of VLBI, then, lies in its ability to make unambiguous
identification of AGN activity. The identification of AGN in lower
resolution radio surveys typically requires detection of excess
radio flux over what is expected from star formation (biasing
samples towards AGN that dominate the star formation), or a dense
gas environment against which the AGN does work and produces
detectable radio lobes (possibly introducing a bias towards AGN in
large groups and clusters). In contrast, VLBI is able to identify
AGN irrespective of the properties of the host galaxy (e.g.
stellar mass and local environment) or the level of star formation
in the system, making it an ideal tool for an exploration of local
AGN and their role in regulating star formation in the nearby
Universe. However, this certainty comes with some limitations; of
the total radio emission generated by AGN activity, only a
fraction will typically be confined to the parsec-scale core/jet
at the site of the central black hole, as extended radio lobes
will often be present. The prominence of the compact core depends
on the source age and orientation (which can result in Doppler
boosting or deboosting of the compact emission). Finally, hotspots
at the site of the jet interaction with the interstellar medium
(as are seen in compact symmetric objects -- CSOs) may also be
compact enough to be visible to VLBI observations, meaning that
although VLBI detections can unambiguously be associated with AGN,
they cannot in every case be associated with AGN {\em cores}.
However, this last case can usually be distinguished on the basis
of morphology.

\begin{figure*}
\begin{minipage}{172mm}
\begin{center}
$\begin{array}{cc}
\includegraphics[width=3.3in]{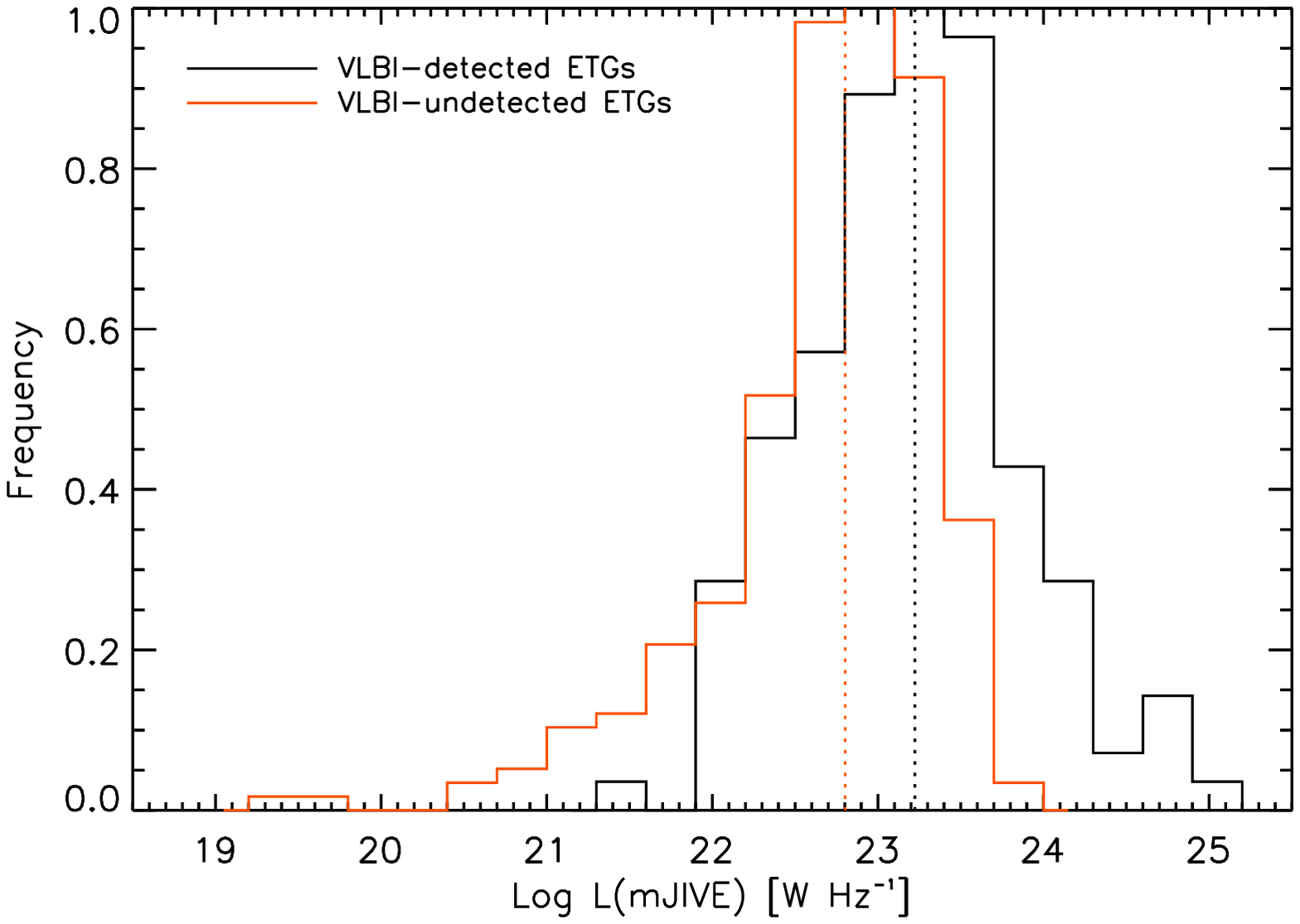} & \includegraphics[width=3.3in]{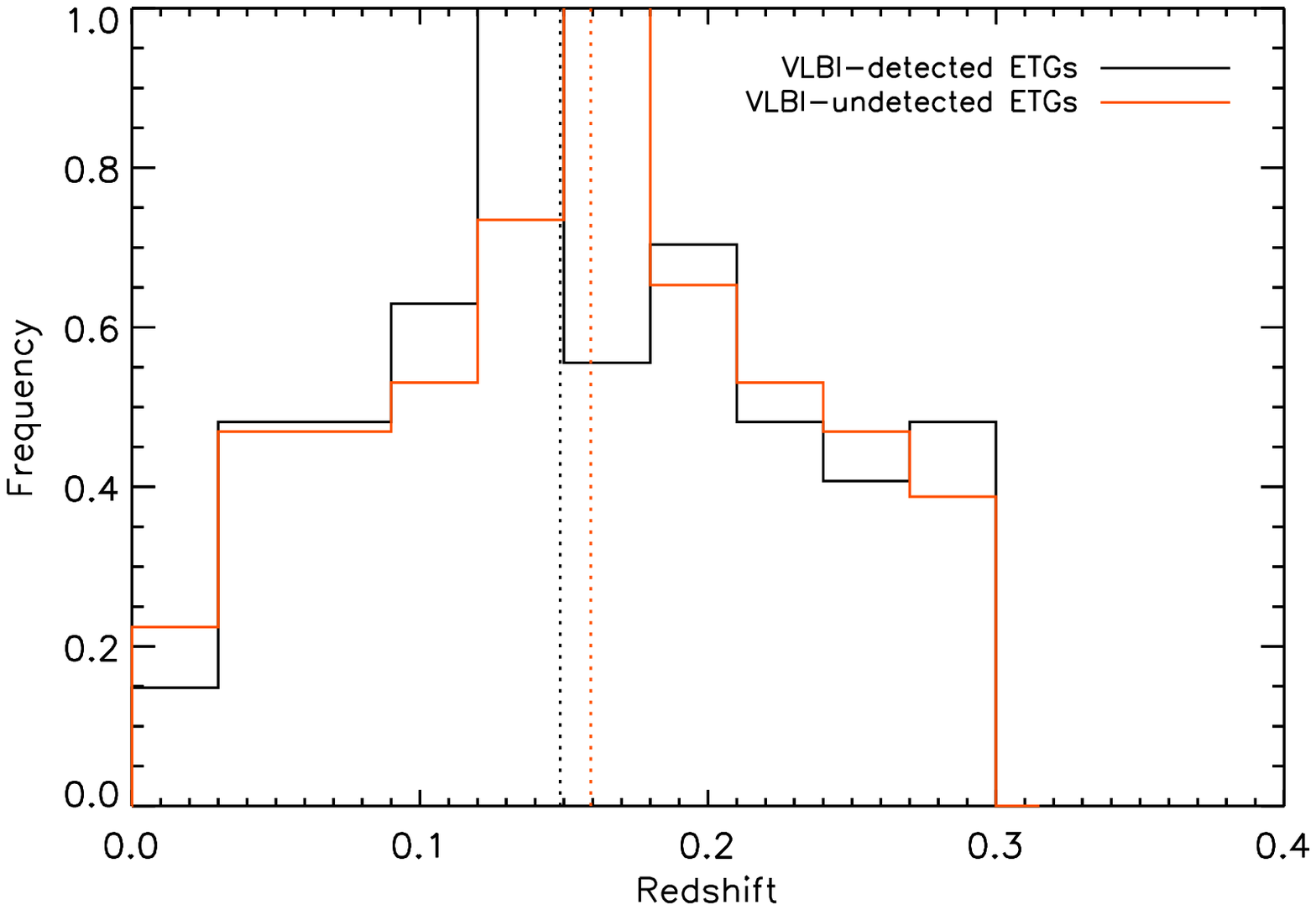}
\end{array}$
\caption{LEFT: mJIVE luminosities of VLBI detections (black) and
upper limits in the VLBI non-detections (orange). RIGHT: The
redshift distribution of the VLBI detections (black) and their
undetected counterparts (orange). Vertical dotted lines indicate
median values.} \label{fig:basic_props}
\end{center}
\end{minipage}
\end{figure*}

The plan for this paper is as follows. In Section 2, we describe
the various datasets that underpin our study. In Section 3, we
explore the processes that trigger the AGN in our galaxy sample.
In Section 4, we study the point in star-formation episodes at
which radio AGN typically appear, and explore the implications for
the regulation of local star formation via AGN feedback. We
summarize our findings in Section 5. Throughout, we employ the
WMAP7 cosmological parameters \citep{Komatsu2011} and photometry
in the AB system \citep{Oke1983}.


\section{Data}

\subsection{Radio VLBI data}
The mJy Imaging VLBA Exploration at 20cm (mJIVE; hereafter) is
using the VLBA to systematically observe objects detected by the
Faint Images of the Radio Sky at Twenty cm (FIRST) survey
\citep{Deller2014}. mJIVE utilizes short segments scheduled in bad
weather or with a reduced number of antennas during which no
highly-rated science projects can be scheduled. After 18 months of
observing, the mJIVE survey has imaged more than 25,000 FIRST
sources, with more than 5000 VLBI detections. While the
sensitivity and resolution varies between mJIVE fields, the median
detection threshold of 1.2 mJy/beam and typical beam size of
6$\times$17 milliarcseconds corresponds to a detection sensitivity
of $\sim10^7$ K, with a typical variation between fields of around
a factor of 2. We refer readers to \citet{Deller2014} for further
details of the survey.


\subsection{SDSS and GALEX photometry}
The mJIVE targets are cross-matched with the latest versions of
the SDSS and GALEX surveys \citep{Martin2005,Morrissey2007}. SDSS
provides five-band ($ugriz$) optical photometry, while GALEX
provides ultraviolet (UV) photometry in two passbands shortward of
3000\AA. For the purposes of this work, we only use the GALEX
near-ultraviolet (NUV) filter, which is centred at 2300\AA.
Following \citet{Shabala2008}, we use a matching radius of 2
arcseconds for the radio-optical matching, which yields high
(96\%) completeness and low (0.3\%) contamination. GALEX
observations are then matched to the SDSS objects with a matching
radius of 4 arcseconds (see e.g. Kaviraj et al. 2007b). Magnitudes
are corrected for Galactic extinction using \citet{Schlegel1998}
and K-corrected using the public \texttt{KCORRECT} code of
\citet{Blanton2007}.


\subsection{Stellar masses and optical AGN diagnostics}
In our analysis below, we employ published stellar masses and
emission-line diagnostics from the latest version of the
publicly-available MPA-JHU value-added SDSS catalogue
\citep[][]{Kauffmann2003,Brinchmann2004,Tremonti2004}\footnote{http://www.mpa-garching.mpg.de/SDSS/DR7/}.
The emission-line class of the galaxy is derived via a standard
line-ratio analysis \citep[][see also Baldwin et al. 1981,
Veilleux et al. 1987, Kewley et al. 2006]{Kauffmann2003}, using
the values of [NII]/H$\alpha$ and [OIII]/H$\beta$ measured from
the SDSS spectra of individual galaxies. Objects in which all four
emission lines are detected with a signal-to-noise ratio greater
than 3 are classified as either `star-forming', `composite',
`Seyfert' or `LINER', depending on their location in the
[NII]/H$\alpha$ vs. [OIII]/H$\beta$ diagram. Galaxies which do not
have a detection in all four lines are classified as
`quiescent' \citep{Kauffmann2003}. We refer readers to
\citet{Brinchmann2004} and \citet{Tremonti2004} for full details
of the modelling.


\subsection{Local environment}
We study the local environments of our galaxies by using the group
catalogue of \citet{Yang2007}, who use a halo-based group finder
to separate the SDSS into over 300,000 structures
{\color{black}with a broad dynamic range}, {\color{black}from}
rich clusters to isolated galaxies. This catalogue provides
estimates of the masses of the host dark matter (DM) haloes of
individual SDSS galaxies, which are related to the traditional
classifications of environment (`field', {\color{black}`group'}
and {\color{black}`cluster'}). Cluster-sized haloes typically have
masses greater than $10^{14}$ M$_{\odot}$, while group-sized
haloes have masses between $10^{13}$ and $10^{14}$ M$_{\odot}$.
Smaller DM haloes constitute what is commonly termed the field
\citep[e.g.][]{Binney1987}.


\subsection{Recent star formation histories}
We estimate parameters governing the recent star formation history
(SFH) of individual galaxies by comparing their GALEX/SDSS $(NUV,
u, g, r, i, z)$ photometry to a library of synthetic photometry,
generated using a large collection of model SFHs. Since VLBI
detections are overwhelmingly found in ETGs (see \S2.7 below), our
SFHs are tailored to these systems. Given that the bulk of the
stellar mass in ETGs forms rapidly at high redshift
\citep[e.g.][]{Trager2000a}, we model the underlying stellar
population using an instantaneous burst at high redshift that
takes places at $z=3$. The recent star formation episode is
modelled using a second instantaneous burst, which is allowed to
vary in age between 0.001 Gyrs and look-back time corresponding to
$z=3$ in the rest-frame of the galaxy, and in mass fraction
between 0 and 1. Our parametrization is similar to previous work
that has quantified early-type SFHs using UV/optical data
\citep[e.g.][Kaviraj et al. 2007b]{Jeong2007}.

To build the library of synthetic photometry, each model SFH is
combined with a metallicity in the range 0.1Z$_{\odot}$ to
2.5Z$_{\odot}$ and a value of dust extinction parametrized by
E$_{\textnormal{B-V}}$ in the range 0 to 0.5, via the empirical
dust prescription of \citet{Calzetti2000}. Photometric predictions
are generated by combining each model SFH with the chosen
metallicity and E$_{\textnormal{B-V}}$ values and convolving with
the stellar models of \citet{Yi2003} through the GALEX and SDSS
filtercurves. Since our galaxy sample spans a range of redshifts,
equivalent libraries are constructed at redshift intervals of
$\delta z=0.02$.

Values of the free parameters are estimated by comparing each
galaxy to every model in the synthetic library and calculating the
model likelihoods ($\exp -\chi^2/2$, e.g. \citet{Sivia1996}). From
the joint probability distribution, each parameter is marginalized
to extract its one-dimensional probability density function (PDF).
The median of this PDF is taken to be the best estimate of the
parameter in question and the 16 and 84 percentile values are used
to calculate an associated `one-sigma' uncertainty. In the
analysis that follows, we explore the median values of the age of
the recent starburst ($t_2$).


\subsection{Visually classified galaxy morphologies}
We use direct visual inspection of SDSS colour images to classify
our galaxies into two broad morphological types: early-type
galaxies (ETGs) and late-type galaxies (LTGs). While
classification into fine morphological classes (e.g. ellipticals,
S0, Sa etc.) is difficult to achieve beyond $z\sim0.2$ using SDSS
images \citep[see e.g.][]{Kaviraj2009a}, splitting our galaxies
into such broad morphological classes is robust to $z\sim0.3$, as
indicated by Figures \ref{fig:etg_examples} and
\ref{fig:ltg_examples} which show typical examples of ETGs and
LTGs across the redshift range studied in this paper.

\begin{figure}
$\begin{array}{c}
\includegraphics[width=3.3in]{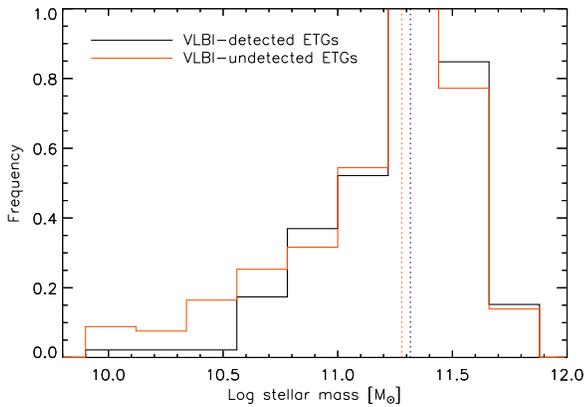}\\
\end{array}$
\caption{The stellar mass distribution of VLBI-detected (black)
and undetected (orange) ETGs in our sample. Vertical dotted lines
indicate median values.} \label{fig:mass_hist}
\end{figure}


\subsection{The final galaxy sample and basic properties}
The final galaxy sample that underpins this study comprises 346
mJIVE targets which have stellar masses and SFRs from the MPA-JHU
catalogue, environments from the Yang et al. group catalogue, star
formation histories derived via SED fitting and
visually-classified morphologies. ETGs and LTGs comprise 76\% (263
galaxies) and 24\% (83 galaxies) of the mJIVE targets
respectively. GALEX coverage is available for around two-thirds of
the sample, so parts of our subsequent analysis that rely on GALEX
data are restricted to this slightly smaller fraction of objects.

We find that the VLBI detection rate in mJIVE targets that are
ETGs is $\sim30$\%, while the corresponding rate in LTGs is
$\sim$3\%. However, since black-hole mass scales with galaxy
stellar mass \citep[e.g.][]{McConnell2011} and ETGs are on average
more massive than LTGs, a fairer comparison can be made by
comparing the LTG detection rate to a subsample of ETGs that have
the same stellar mass and redshift distribution as the LTGs. The
detection rate in this matched ETG subsample is $\sim$26\%. Note
that the distributions of FIRST fluxes of the LTGs and the ETG
subsample are very similar, making these values a reasonable
estimate of the relative detection rate in the two morphological
classes. Radio AGN are, therefore, almost an order of magnitude
more frequent in ETGs than in their LTG counterparts, in agreement
with the past literature \citep[e.g.][]{Ledlow2001}. The
VLBI-detected LTGs will be the subject of a forthcoming paper
(Kaviraj et al., in prep). Given the dearth of radio AGN in the
LTGs, our subsequent analysis focuses only on the ETG population.

\begin{figure}
$\begin{array}{c}
\includegraphics[width=3.3in]{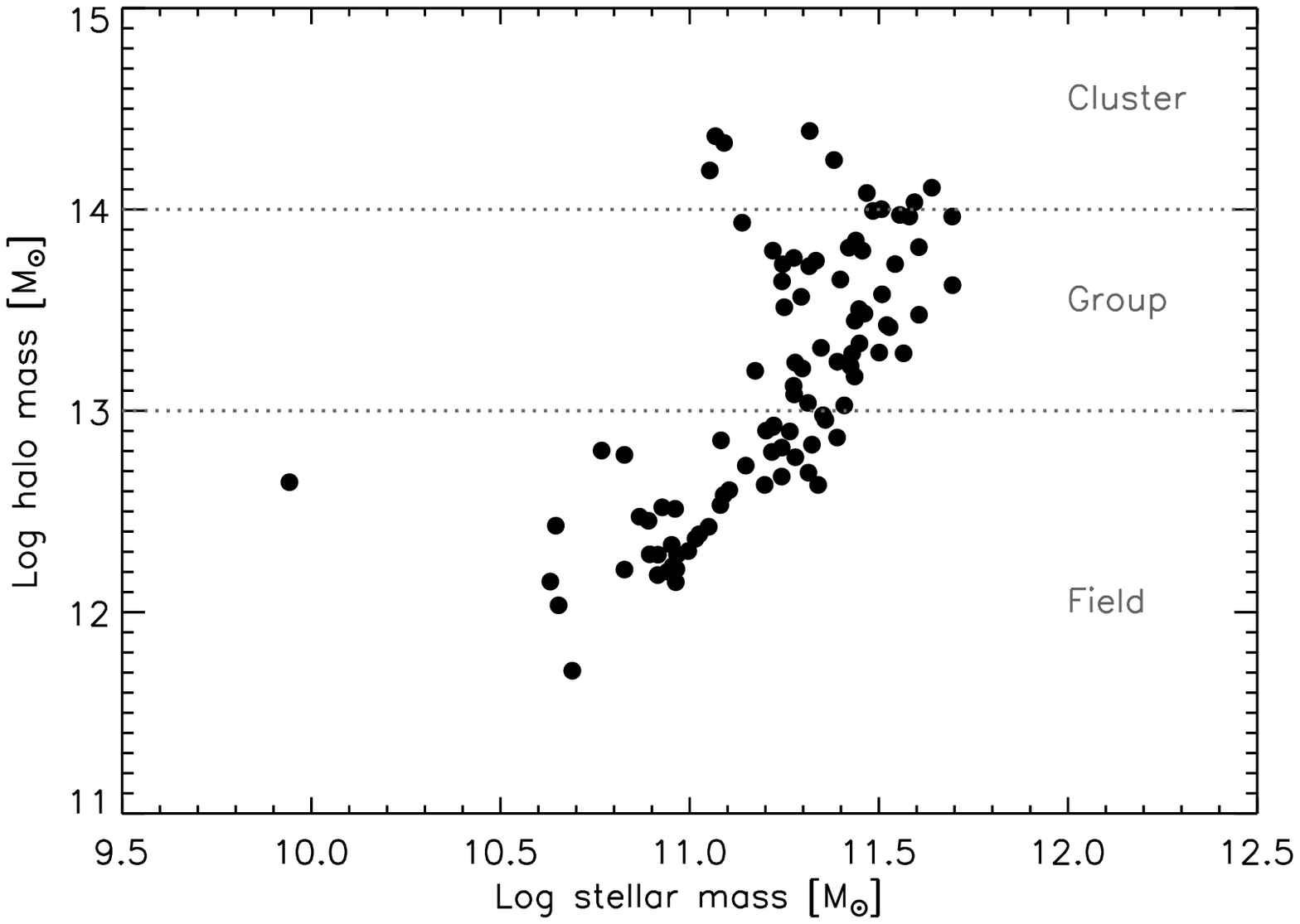}\\
\includegraphics[width=3.3in]{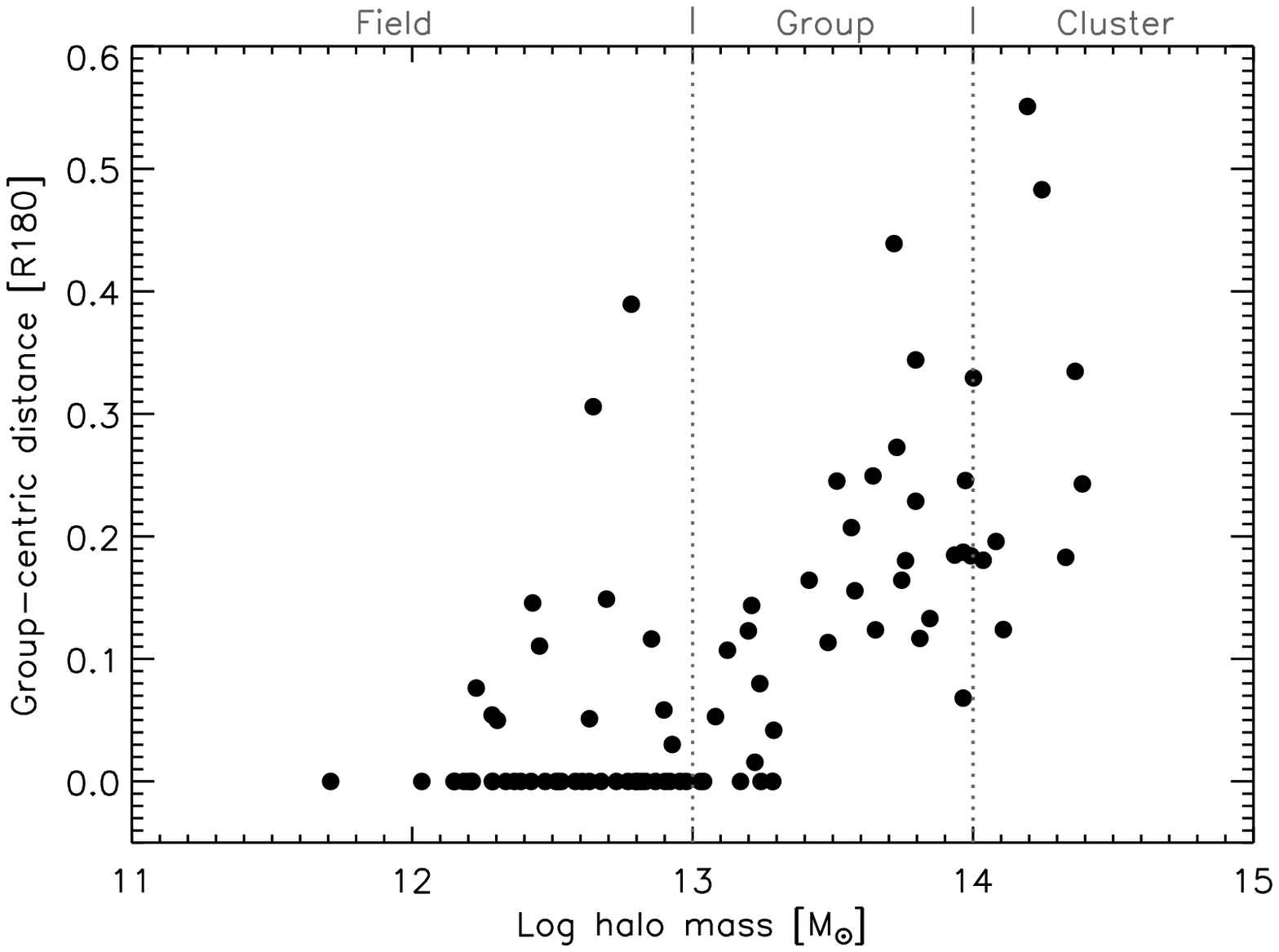}
\end{array}$
\caption{Dark matter halo vs stellar mass (top) and group-centric
radius vs halo mass (bottom) for ETGs in our sample. The
\citet{Yang2007} group catalogue define dark matter halos as
having an over-density of 180. The group-centric radii are,
therefore, calculated by dividing the distance of the galaxy in
question by $R_{180}$.}
\label{fig:environment}
\end{figure}

Since the mJIVE targets are, by construction, all detected by
FIRST, it is useful to define a `control' sample of ETGs that are
not FIRST-detected and to which we will compare the ETG VLBI
detections at various points in our analysis below. This control
sample is selected via the Galaxy Zoo project \citep{Lintott2008},
which has used 500,000+ volunteers from the general public to
produce accurate morphological classifications for the entire SDSS
via visual inspection of SDSS colour images. The galaxies in our
ETG control sample have a Galaxy Zoo `early-type' probability of
more than 90\% and are selected to have the same redshift, stellar
mass and environment distribution as the ETGs with VLBI
detections.

Before we begin our analysis, we briefly explore the typical AGN
luminosities being sampled here and whether AGN detectability
varies across our redshift range of interest (possibly leading to
selection effects that might affect our analysis). The left panel
of Figure \ref{fig:basic_props} presents the radio luminosities of
the AGN in our sample. The black histogram indicates the radio
luminosities of our VLBI-detected systems, while the orange
histogram shows upper limits in galaxies that are not detected by
mJIVE. The median radio luminosity of the VLBI detections is
around 10$^{23.5}$ W Hz$^{-1}$, while the lowest luminosities are
around 10$^{22}$ W Hz$^{-1}$ - the AGN studied here are,
therefore, typically brighter than this latter value. The
right-hand panel of this figure shows that the VLBI detections and
non-detections have similar redshift distributions. A
Kolmogorov-Smirnov (KS) test yields a p-value of 0.64, indicating
that the two redshift samples are highly likely to be drawn from
the same parent distribution. The similarity in the redshift
distributions indicates that the detectability of our AGN in VLBI
does not vary in the redshift range studied here ($z<0.3$).


\section{Triggering of local AGN}
We begin by exploring the processes that are likely to be fuelling
the radio AGN in our sample. There are several sources of gas that
may trigger AGN, such as internal stellar mass loss (which feeds
the internal hot gas reservoir), cluster-scale cooling flows
(which will operate in the central galaxies of clusters) and
mergers. Figure \ref{fig:mass_hist} shows that the stellar mass
distribution of the VLBI detections is similar to that of their
undetected counterparts. A KS test between the two samples yields
a p-value of 0.23, indicating that they are likely to be drawn
from the same parent distribution. This similarity suggests that
the AGN are unlikely to be triggered by internal stellar mass
loss, since more massive galaxies will produce larger gas
reservoirs via this process and one might expect the host galaxies
of our VLBI detections to then be preferentially more massive.



Further insight into the gas supply can be gained by studying the
local environments of our AGN. The top panel of Figure
\ref{fig:environment} indicates that the VLBI detections in this
study preferentially lie outside clusters (i.e. in relatively
low-density environments). In addition, those that do inhabit
clusters are not central galaxies but are typically found at
reasonably large group-centric radii (bottom panel of Figure
\ref{fig:environment}). The vast majority of radio AGN in this
particular sample are, therefore, unlikely to be fuelled by
cluster-scale cooling flows.

\begin{figure}
$\begin{array}{c}
\includegraphics[width=3.3in]{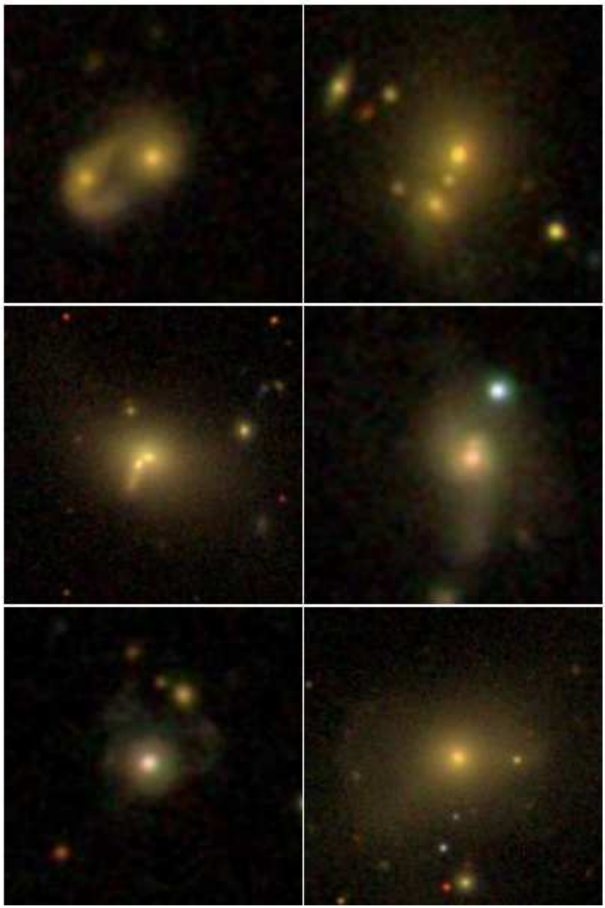}
\end{array}$
\caption{Examples of VLBI-detected ETGs in mergers. The top two
rows indicate systems that are ongoing mergers, while the bottom
row indicates systems that are postmergers.}
\label{fig:merger_examples}
\end{figure}

While the arguments above point towards mergers being an important
trigger for our radio AGN, we quantify this by visually inspecting
our VLBI detections and flagging systems that are either ongoing
mergers (two interacting cores with tidal bridges between them,
see top and middle rows of Figure \ref{fig:merger_examples}) or
postmergers (one nucleus with tidal debris around it, see bottom
row of Figure \ref{fig:merger_examples}). We note that a drawback
of the standard depth (51 second) SDSS images is that faint tidal
features, such as those produced by relatively gas-poor or low
mass-ratio mergers, can be difficult to detect in the images
\citep[e.g.][]{Kaviraj2010}. Hence, the merger fraction calculated
here is strictly a lower limit. Therefore, in addition to
identifying ongoing and post-mergers via visual inspection, we
also search for neighbouring objects around each galaxy that are
within a projected separation of 30 kpc, with either spectroscopic
redshifts with a maximum velocity differential of 500 km s$^{-1}$
or photometric redshifts that are, within their uncertainties,
consistent with the galaxy in question. We restrict these
photometric neighbours to objects that have accurate photometric
redshifts (uncertainties less than 20\%). The 30 kpc separation
employed is motivated by work which shows that star-formation and
AGN activity can be efficiently triggered when the separation
between interacting galaxies is within 30 kpc \citep[see
e.g.][]{Patton2002,Scudder2012,Scott2014}. To quantify the role of
merging in triggering our radio AGN, we perform an identical
exercise (visual inspection + neighbours analysis) on our control
sample of ETGs.

Table 1 presents the merger fractions for the VLBI-detected ETGs
and their control counterparts, split by environment. We provide
merger fractions with and without the neighbour analysis (shown in
brackets). Note that the visually-classified merger fractions in
the control sample are consistent with the findings of past work
which has calculated merger fractions in massive galaxies via
visual inspection of SDSS images \citep{Darg2010a,Darg2010b}. The
bottom row in this table indicates the enhancement of the merger
fraction in the AGN population compared to their control
counterparts. We find that the merger fractions in the AGN
population are several factors higher than that in the general
galaxy population, irrespective of the local environment of the
systems. These results are generally consistent with the broader
literature, in which a coincidence of AGN and merger activity (at
various stages of the merger process) has been reported
\citep[e.g.][]{Schawinski2010,Inskip2010,Ellison2011,Carpineti2012,Koss2010}.
Recalling that our merger fractions are strictly lower limits, it
appears reasonable to conclude that an important trigger for the
radio AGN in our sample is galaxy merging.

We note that, given these high merger fractions and the fact that
our VLBI detections are not central galaxies of clusters, our
subsequent analysis may largely probe `cold-mode' AGN, which are
fuelled via accretion from a cold gas disk fed by mergers and
interactions \citep[e.g.][]{Hardcastle2007,Best2012,Shabala2012},
rather than `hot-mode' AGN, which are fuelled via direct cooling
of hot gas. It is difficult to quantify the relative proportion of
AGN in either fuelling mode because the shallow SDSS images
inevitably lead to an underestimate of the merger fractions and
therefore a corresponding underestimate of the fraction of
cold-mode systems.

\begin{table*}
\begin{minipage}{172mm}
\begin{center}
\caption{Merger fractions for radio AGN and for a control sample
of ETGs, split by environment. The last row indicates the relative
enhancement in the merger fraction in AGN compared to the general
ETG population. The values in brackets include the neighbour
analysis, while those without brackets only include systems that
are flagged as ongoing or post-mergers via visual inspection (see
text in \S3 for details).}
\begin{tabular}{lrr}\hline

                      & Field                                     & Groups and clusters\\\hline
    AGN               & 0.36$^{\pm 0.03}$ (0.45$^{\pm 0.04}$)   & 0.28$^{\pm 0.07}$ (0.33$^{\pm 0.07}$)\\
    Control           & 0.11$^{\pm 0.02}$ (0.14$^{\pm 0.02}$)   & 0.06$^{\pm 0.01}$ (0.08$^{\pm 0.01}$)\\ \hline
    f(AGN)/f(Control) & 3.3  (3.2)                                & 4.7  (4.1) \\

\end{tabular}
\end{center}
\label{tab:merger_fractions}
\end{minipage}
\end{table*}

\begin{figure}
$\begin{array}{c}
\includegraphics[width=3.3in]{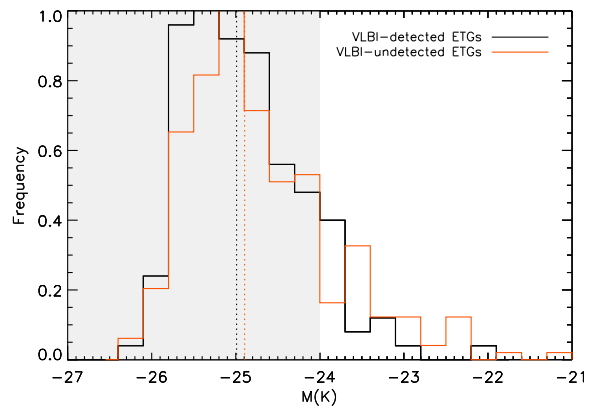}\\
\end{array}$
\caption{Absolute K-band luminosity distribution of the ETGs in
our sample. The shaded region indicates the luminosity range in
which AGN feedback is expected to play an important role in
regulating star formation \citep[e.g.][]{Benson2003}. Vertical
dotted lines indicate median values.} \label{fig:M_K}
\end{figure}


\section{Time delay in the onset of AGN and implications for
regulation of star formation} We proceed by exploring the point in
the evolution of star formation episodes when radio AGN typically
appear, which has important implications for the potential impact
of AGN feedback on star formation. For feedback to be most
effective, the AGN must appear promptly after the onset of star
formation (or at least before the star formation peaks), when the
systems are gas-rich. However, if the onset of the AGN takes place
several dynamical timescales into the starburst, then a large
fraction of the gas reservoir will have been depleted before the
AGN has a chance to interact with it, reducing its role in
regulating star formation. It is worth noting here that the ETGs
in our study (see Figure \ref{fig:M_K}) are in the luminosity
range where AGN feedback is expected to play a key role in
regulating and quenching star formation
\citep[e.g.][]{Benson2003}.

\begin{figure}
$\begin{array}{c}
\includegraphics[width=3.3in]{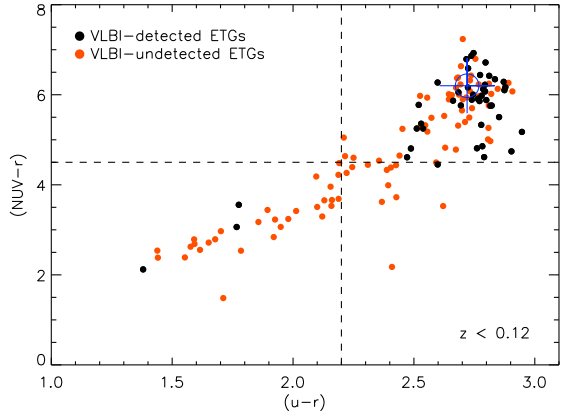}
\end{array}$
\caption{UV-optical colours of ETGs in our sample. The
VLBI-detected ETGs are shown in black, while ETGs that are
undetected are shown in orange. The median colour of the general
ETG population, from the SDSS control sample of ETGs, is shown
using the blue crosshair. The dashed lines demarcate the blue
cloud (bottom left region) and red sequence (top right region), as
indicated by the literature \citep[see
e.g.][]{Strateva2001,Wyder2007}.} \label{fig:colours}
\end{figure}

Figure \ref{fig:colours} shows that the VLBI-detected ETGs
primarily reside on the UV-optical red sequence. We restrict this
analysis to $z<0.12$ (135 galaxies), because the GALEX UV red
sequence detection rate is $\sim$95\% in this redshift range but
drops rapidly thereafter (see Figure 1 in Kaviraj et al. 2007b).
The high sensitivity of the UV wavelengths to star formation means
that even small mass fractions of young stars (of the order of a
percent or less) will drive ETGs into the UV-optical blue cloud
\citep[e.g.][Kaviraj et al. 2007b]{Yi2005}, making the UV-optical
colour space the most effective route to separating star-forming
systems from their more quiescent counterparts. Figure
\ref{fig:colours} therefore suggests that radio AGN typically
inhabit galaxies in which star formation activity is low i.e. has
declined significantly since the onset of the starburst. Table 2
presents the blue-cloud to red-sequence ratios for the
VLBI-detected ETGs and the control sample, where galaxies that
have $(NUV-r)<4.5$ are classified as `blue' and those with
$(NUV-r)>4.5$ are classified as `red' \citep{Wyder2007}.  The
blue-to-red ratios in the VLBI-detected ETGs and the control
sample are 9\% and 25\% respectively (the VLBI-undetected ETGs
show a ratio of 80\%). The likelihood of finding VLBI detections
in the blue cloud is, therefore, around a factor of 3 lower than
in the general ETG population.

If AGN were triggered promptly after the onset of star formation,
one would expect the blue-to-red ratio in the VLBI-detections to
be higher than that in the general ETG population (the opposite to
what is observed). It is worth exploring whether our red-sequence
AGN could originally have been triggered promptly while their host
galaxies were in the blue cloud and have remained active as their
host galaxies transited across the colour-colour diagram. This is
possible if AGN lifetimes are comparable to the time it takes for
a galaxy to transit from the blue cloud to the red sequence. The
typical lifetimes of AGN (i.e. the lifetime of the `on' phase), in
galaxies within our mass range of interest, are a few $10^7$ yr
\citep[e.g.][see their Figure 13]{Shabala2008}. We can estimate a
lower limit on blue-to-red transit times, via past work on nearby
post-starburst (E+A) galaxies. E+A galaxies are systems that do
not show emission lines that trace ongoing star formation (e.g.
H$\alpha$) but exhibit deep Balmer absorption lines that are
indicative of significant recent star formation. These galaxies
have, therefore, experienced a recent starburst which has been
truncated. Kaviraj et al. (2007c) have shown that the median
blue-to-red transit time of E+A galaxies in the nearby Universe is
$\sim1.5$ Gyr (see their Figure 7), consistent with theoretical
work on the likely timescales for this transit (e.g. Kaviraj et
al. 2011a). Since star formation has been completely quenched in
these systems, E+A galaxies will transit from blue cloud to red
sequence in the \emph{shortest} possible time.

Given that typical AGN lifetimes are much shorter than $\sim$1.5
Gyr, the radio AGN that are observed on the UV-optical red
sequence are unlikely to have appeared in their host galaxies when
these systems were originally in the blue cloud. Even if the
observed AGN are the result of re-triggering on timescales of
around a Gyr \citep[e.g.][]{Ensslin2001}, the paucity of AGN in
the blue cloud and the arguments above indicate that the AGN
episodes that directly follow the onset of starbursts in ETGs are
unlikely to be prompt.

It is instructive to estimate the magnitude of the time delay
between the onset of star formation and the onset of the AGN, by
comparing the age of the starburst (estimated in Section 2.4 via
SED fitting) to the dynamical timescale ($\tau_{dyn}$) in
individual galaxies, defined as\footnote{Using the equation of
motion $s=\frac{1}{2}at^2$, the dynamical timescale follows by
setting $a=g=(GM/R^2)$ and $s=R$, where $M$ and $R$ are the mass
and radius of the system respectively.}:

\begin{equation}
\tau_{dyn} = \Big(\frac{2R^3}{GM}\Big)^{1/2}.
\end{equation}

\begin{figure}
$\begin{array}{c}
\includegraphics[width=3.3in]{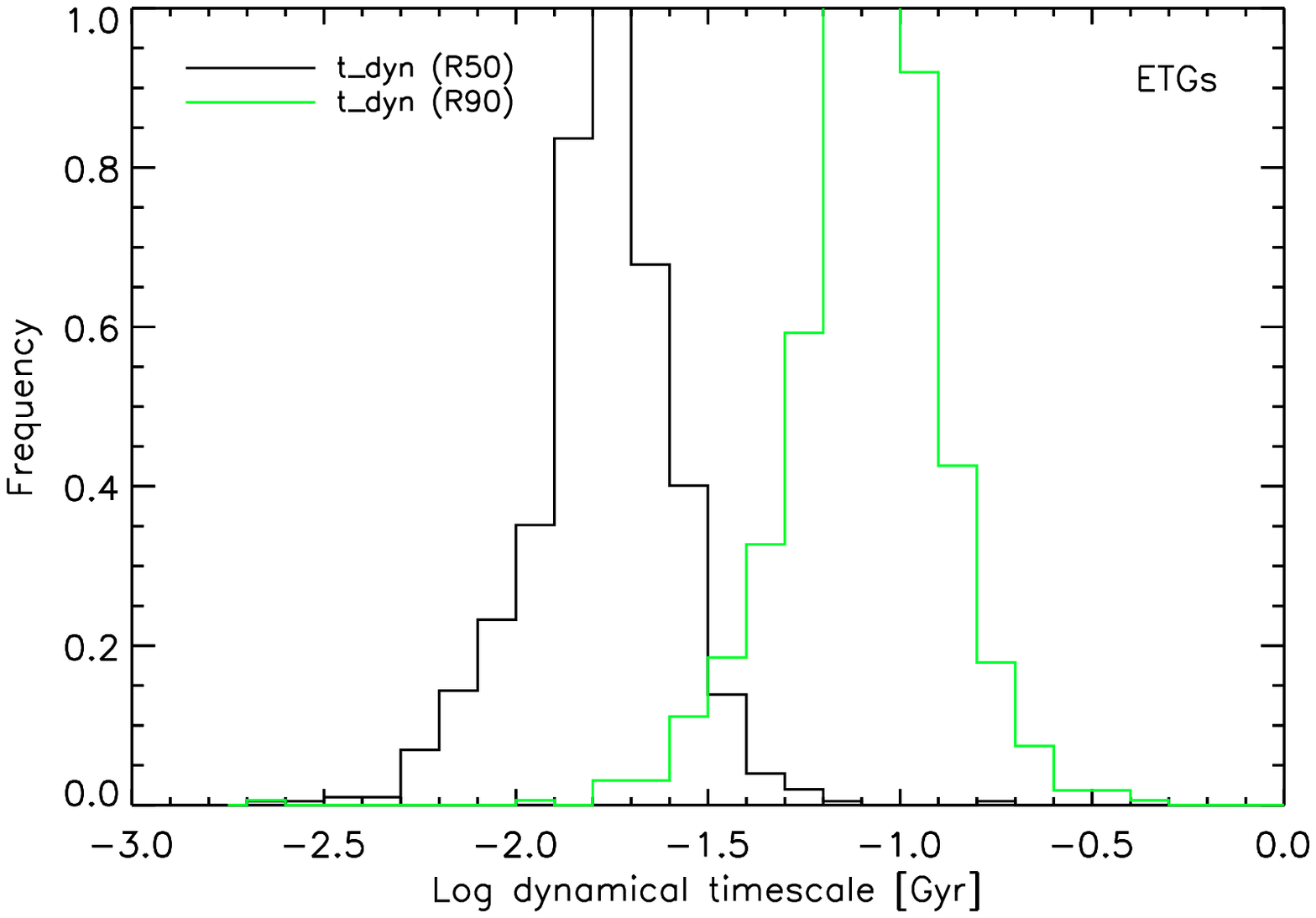}\\
\includegraphics[width=3.3in]{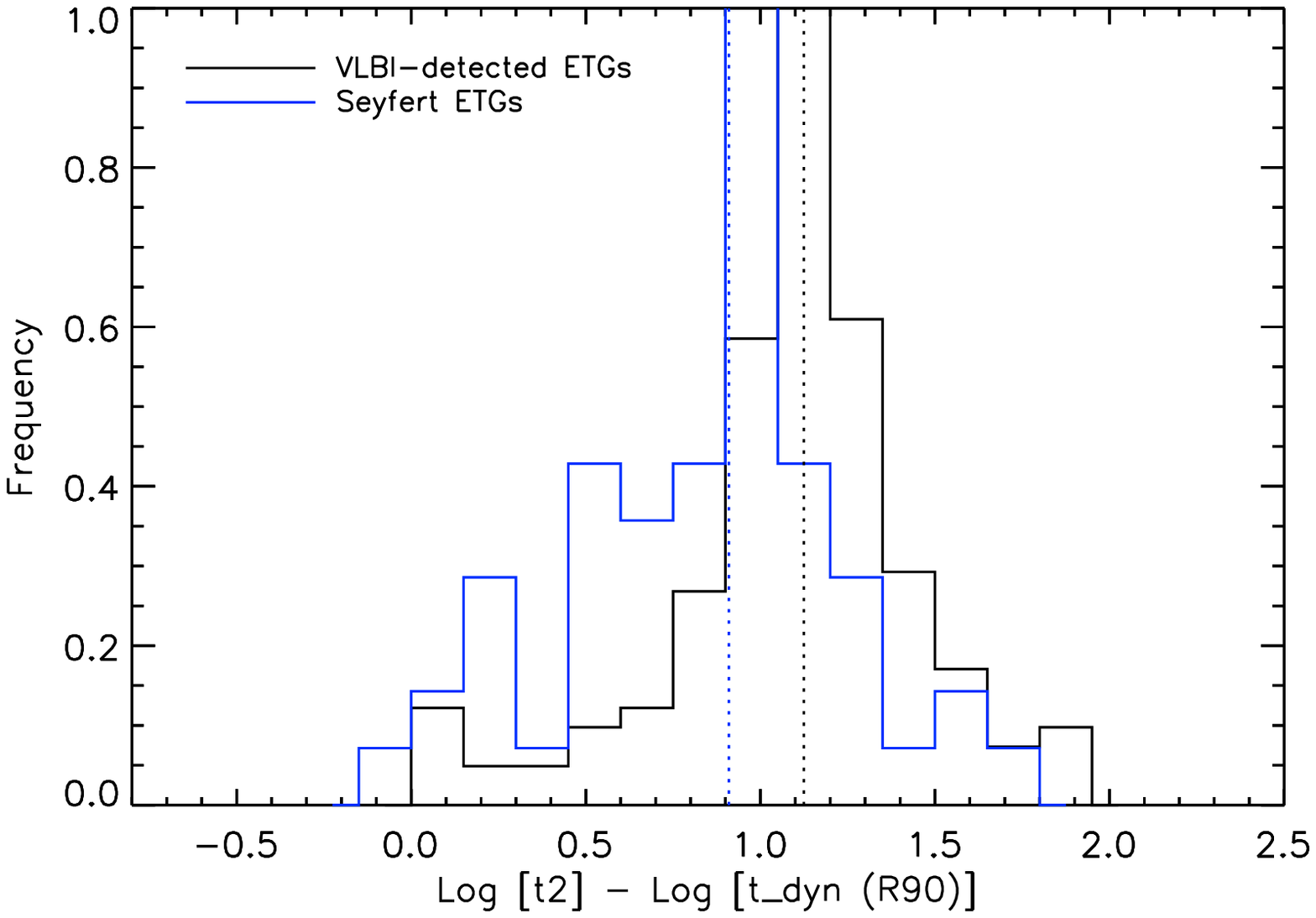}
\end{array}$
\caption{TOP: Dynamical timescales of the ETGs in our sample,
calculated both using radii that contain 50\% (R$_{50}$; black
histogram) and 90\% (R$_{90}$; blue histogram) of the $r$-band
Petrosian flux as the value for $R$ in Eqn 1. BOTTOM: The ratio of
the age of the starburst and the dynamical timescale in
VLBI-detected (black histogram) and Seyfert (blue histogram) ETGs.
Vertical dotted lines indicate median values.} \label{fig:tdyn}
\end{figure}

In the top panel of Figure \ref{fig:tdyn}, we present the
$\tau_{dyn}$ values of our ETGs, calculated both using radii that
contain 50\% (R$_{50}$; black histogram) and 90\% (R$_{90}$; blue
histogram) of the $r$-band Petrosian flux as the value for $R$ in
Eqn 1. While our $\tau_{dyn}$ values have been calculated using
the stellar properties of the galaxies in question, our values
using R$_{90}$ are very similar to the star-formation timescales
of gas disks in local ETGs, calculated via CO measurements
\citep[0.05-0.2 Gyrs;][]{Young2002}. In the bottom panel of Figure
\ref{fig:tdyn}, we present the ratio of the age of starburst and
the dynamical timescale in our VLBI detections (calculated using
R$_{90}$). Dividing the starburst age by the dynamical timescale
provides a normalized quantity, enabling a fairer comparison of
galaxies that have different radii and masses. We find that the
ages of the star formation episodes in our VLBI detections are
typically several multiples of the dynamical timescales of the
galaxies in question, suggesting that these radio AGN are
triggered relatively late into the associated starbursts (note
that the figure indicates that this is also true of Seyferts ETGs,
in line with the findings of the recent literature). In such a
scenario, the ability of the AGN to regulate star formation will
be reduced, since a significant fraction of the gas reservoir is
likely to have been depleted before the AGN switches on. While the
AGN may well couple to the remaining gas reservoir and drive an
outflow, the overall impact of such feedback episodes will be to
mop up residual gas at a point where star formation has already
declined.

\begin{table}
\begin{center}
\caption{Ratio of blue-cloud to red-sequence objects in our ETGs.
Only ETGs at $z<0.12$ are considered here because the UV red
sequence detection rate is $\sim$95\% in this redshift range but
drops rapidly thereafter.}
\begin{tabular}{lc}\hline

                         & Blue-to-red ratio\\\hline
    VLBI-detected ETGs   & 0.09$^{\pm 0.04}$\\
    Control sample ETGs  & 0.25$^{\pm 0.06}$\\
    VLBI-undetected ETGs & 0.80$^{\pm 0.15}$
\end{tabular}
\end{center}
\label{tab:merger_fractions}
\end{table}

Our results are consistent with recent observational and
theoretical work which suggests that there is a time delay between
the onset of star formation episodes and the onset of the
associated AGN activity in the nearby Universe. As noted in the
introduction, a number of studies have shown that the peak of
optically-identified AGN activity is delayed compared to the onset
of star formation by several hundred Myr. This is found to be the
case across the full spectrum of star formation activity, from
strongly star-forming systems such as luminous infrared galaxies
\citep{Kaviraj2009a} to the more weakly star-forming ETGs
\citep[e.g.][]{Davies2007,Schawinski2007,Wild2010,Shabala2012,Yesuf2014}.

Some physical explanations have been proposed in the recent
literature to explain this observed time lag. Supernova feedback
during the peak phase of star formation activity may disrupt the
supply of gas onto the black hole, with preferential fuelling in
the post-starburst phase from stellar mass loss \citep{Wild2010}.
It remains unclear if fuelling purely via stellar mass loss is
appropriate for systems that are likely to be accreting material
in the cold phase
\citep[e.g.][]{Lonsdale2003,Kondratko2006,Nenkova2008}. The cause
of this delay could also be dynamical in nature. During the peak
of the starburst, when the system is rich in gas, most regions are
unstable and more conducive to star formation, making this the
favoured mode of gas consumption \citep{Cen2012}. Inflow on to the
black hole requires further removal of angular momentum, which
becomes efficient only after most of the gas has been consumed by
star formation activity \citep{Hopkins2011}, introducing a time
lag between the onset of star formation and the triggering of the
AGN. Realistic hydro-dynamical models appear to naturally produce
a time lag that is qualitatively similar to what is observed in
nearby ETGs (Hopkins 2011, see also Cox et al. 2006, Robertson et
al. 2006).

It should be noted that our results do not imply that AGN feedback
does not take place at all in the nearby Universe. As noted in the
introduction, strong evidence exists for the regulation of star
formation in central cluster galaxies via AGN-driven heating of
cooling flows. However, while the AGN appears to couple
efficiently to gas that is acquired via cooling, the onset of the
AGN appears to be significantly delayed when the gas is brought in
via mergers. This time lag implies that the impact of AGN on star
formation driven by mergers is relatively limited,
because they are triggered at a point in the star formation
episode when the gas reservoir has already been significantly
depleted.

The results of this study are strongly aligned with work on star
formation in ETGs. The recent literature that leverages data in
the UV wavelengths has demonstrated that, contrary to our
classical notion of ETGs being passively-evolving galaxies,
widespread star formation exists in these galaxies (e.g. Yi et al.
2005; Kaviraj et al. 2007b; Jeong et al. 2007). A strong
correspondence is observed between blue UV colours and
morphological disturbances, indicating that the star formation in
merger-driven. However, the major merger rate is too low to
reproduce the fraction of disturbed ETGs, implying that minor
mergers drive the stellar mass growth in these systems (Kaviraj et
al. 2009, 2011b), with a significant minority (20-30\%) of the
stellar mass forming after $z \sim 1$ (Kaviraj et al. 2008; see
also Kaviraj 2014b). Based on our findings here, it is likely that
the merger-driven star formation that is observed in nearby ETGs
is possible precisely because the associated AGN activity is
triggered late into these star formation episodes and cannot
strongly regulate or quench the starburst. The weak coupling of
AGN to merger-driven star formation therefore makes this the
dominant mode of stellar mass growth in nearby ETGs.

What do our results imply for our general paradigm of
AGN-regulated galaxy growth? Two issues are worth noting here.
First (as mentioned above), since our sample is drawn from the
general galaxy population, it is not dominated by central cluster
galaxies and does not offer constraints on the role of AGN in such
environments. The past literature, however, provides compelling
evidence for feedback in such environments, through AGN heating of
cooling flows. In the nearby Universe strong AGN feedback is
therefore likely to be most efficient in (and largely restricted
to) these environments. Second, since the galaxies studied here
are local, our results do not offer insight into the interaction
of AGN with their host galaxies at high redshift. Given that the
cosmic star formation and black hole accretion rates peak at
$z\sim2$, it is around these redshifts that the bulk of today's
stellar mass formed. This is also the epoch at which the observed
scaling relations -- such as the M$_{\textnormal{BH}}$-$\sigma$
correlation -- were established, most likely over short timescales
($<1$ Gyr), as indicated by the high [$\alpha$/Fe] ratios in
massive ETGs \citep[e.g.][]{Trager2000a,Thomas2005}.

For this high-redshift scenario to be strongly influenced by AGN
feedback, the situation must be different at high redshift. In
particular, AGN must be triggered in `normal' (i.e. non-merging)
LTGs, which dominate the star formation budget at these epochs
\citep[e.g.][]{Kaviraj2013a} and couple efficiently to their host
galaxies in order to produce the scaling relations over short
timescales. The latter requires prompt triggering of the AGN,
presumably leading to a smaller (or negligible) colour offsets
between the star forming and AGN populations at these epochs.
While the scope of this study does not allow us to probe these
issues (since our sample is local), in forthcoming work we will
study the co-evolution of AGN and their host galaxies at higher
redshift, and present preliminary evidence that the requirements
mentioned above are indeed satisfied by the radio AGN population
around the epoch of peak cosmic star formation.


\section{Summary}
We have studied $\sim$350 massive, nearby galaxies, observed by
the mJIVE survey, to explore the role of AGN in regulating star
formation in nearby galaxies. The $\gtrsim 10^7$ K temperatures
required for an mJIVE detection cannot be achieved via star
formation alone, allowing us to unambiguously detect radio AGN and
study their role in galaxy evolution in the nearby Universe.

VLBI detections are overwhelmingly found in ETGs, with the
detection rate in LTGs being an order of magnitude lower. The
VLBI-detected ETGs in this study are not preferentially more
massive than their undetected counterparts and are typically drawn
from relatively low-density environments (i.e. are not the central
galaxies of clusters). They also show significantly higher merger
fractions than the general ETG population. Taken together, this
suggests that a significant trigger for nearby radio AGN is galaxy
merging.

Our analysis indicates that the onset of merger-triggered radio
AGN is typically not prompt and takes place several dynamical
timescales into the starburst event, making it unlikely that the
AGN provides strong regulation of this star formation. While the
AGN may well couple to the remaining gas reservoir, the overall
impact of this process will be to mop up residual gas, in a system
where the gas reservoir has already been significantly depleted.

While the wider literature provides strong evidence for AGN
feedback in systems where the black hole is fuelled by the cooling
of hot gas (e.g. in central cluster galaxies), our results
indicate that this is not the case when the gas is introduced into
the system via a merger, at least at low redshift. The inability
of the black hole to couple strongly to accreted gas then makes
galaxy merging the dominant mode of stellar mass growth in nearby
ETGs, precisely because the AGN is not able to regulate (and
rapidly quench) merger-driven star formation. This scenario agrees
with the recent literature which demonstrates widespread star
formation in ETGs, which is indeed driven by minor mergers.

Finally, since the galaxies studied here are local, our results do
not offer insight into the interaction of AGN with their host
galaxies at high redshift. Given that the cosmic star formation
and black hole accretion rates peak at $z\sim2$, it is around
these redshifts that the bulk of today's stellar mass must have
formed and galaxy scaling relations, such as the
M$_{\textnormal{BH}}$-$\sigma$ correlation, largely put in place.
As noted in \S4, for these scaling relations to be established at
high redshift over short timescales, one requires a qualitatively
different interaction between AGN and their hosts at these epochs.
AGN in the early Universe must operate in normal LTGs and be
triggered promptly in order to provide efficient regulation of
star formation. While our local sample does not offer insight into
AGN-galaxy co-evolution at these epochs, similar studies at
high-redshift are keenly anticipated in order to probe the role of
AGN in stellar mass growth around the epoch of peak cosmic star
formation.


\section*{Acknowledgements}
We thank Martin Hardcastle, Andrew King, Rachel Somerville, Frank
van den Bosch, Meg Urry and Kevin Schawinski for interesting
discussions. SK is grateful for support from the University of
Tasmania (UTas) via a UTas Visiting Scholarship and acknowledges a
Senior Research Fellowship from Worcester College Oxford. SSS
acknowledges an ARC Early-Career Fellowship (DE130101399). ATD was
supported by an NWO Veni Fellowship.


\nocite{Baldwin1981,Veilleux1987,Kewley2006,Kaviraj2008b,Trager2000,Cox2006,Robertson2006,Rupke2010,Torrey2011,Kaviraj2007a,Kaviraj2007c,Kaviraj2007b,Kaviraj2011,Kaviraj2011c,Kaviraj2014a,Kaviraj2014b,Kaviraj2008}


\bibliographystyle{mn2e}
\bibliography{references}

\begin{thebibliography}{}

\bibitem[\protect\citeauthoryear{{Abazajian}, {} \& {et al.}}{{Abazajian}
  et~al.}{2009}]{Abazajian2009}
{Abazajian} K.~N.,  {}   {et al.} 2009, \apjs, 182, 543

\bibitem[\protect\citeauthoryear{{Alexander}, {Swinbank}, {Smail}, {McDermid}
  \& {Nesvadba}}{{Alexander} et~al.}{2010}]{Alexander2010}
{Alexander} D.~M.,  {Swinbank} A.~M.,  {Smail} I.,  {McDermid} R.,
  {Nesvadba} N.~P.~H.,  2010, \mnras, 402, 2211

\bibitem[\protect\citeauthoryear{{Baan}}{{Baan}}{2007}]{Baan2007}
{Baan} W.~A.,  2007, in {Chapman} J.~M.,  {Baan} W.~A.,  eds, IAU Symposium
  Vol.~242 of IAU Symposium, {Arp 220 IC 4553/4: understanding the system and
  diagnosing the ISM}.
pp 437--445

\bibitem[\protect\citeauthoryear{{Baldwin}, {Phillips} \&
  {Terlevich}}{{Baldwin} et~al.}{1981}]{Baldwin1981}
{Baldwin} J.~A.,  {Phillips} M.~M.,    {Terlevich} R.,  1981, PASP, 93, 5

\bibitem[\protect\citeauthoryear{{Benson}, {Bower}, {Frenk}, {Lacey}, {Baugh}
  \& {Cole}}{{Benson} et~al.}{2003}]{Benson2003}
{Benson} A.~J.,  {Bower} R.~G.,  {Frenk} C.~S.,  {Lacey} C.~G.,  {Baugh} C.~M.,
     {Cole} S.,  2003, ApJ, 599, 38

\bibitem[\protect\citeauthoryear{{Best} \& {Heckman}}{{Best} \&
  {Heckman}}{2012}]{Best2012}
{Best} P.~N.,  {Heckman} T.~M.,  2012, \mnras, 421, 1569

\bibitem[\protect\citeauthoryear{{Best}, {Kauffmann}, {Heckman} \&
  {Ivezi{\'c}}}{{Best} et~al.}{2005}]{Best2005}
{Best} P.~N.,  {Kauffmann} G.,  {Heckman} T.~M.,    {Ivezi{\'c}} {\v Z}.,
  2005, \mnras, 362, 9

\bibitem[\protect\citeauthoryear{Binney \& Tremaine}{Binney \&
  Tremaine}{1987}]{Binney1987}
Binney J.,  Tremaine S.,  1987, Galactic Dynamics, first edn.
Princeton Series in Astrophysics, Princeton University Press

\bibitem[\protect\citeauthoryear{{Blanton} \& {Roweis}}{{Blanton} \&
  {Roweis}}{2007}]{Blanton2007}
{Blanton} M.~R.,  {Roweis} S.,  2007, \aj, 133, 734

\bibitem[\protect\citeauthoryear{{Brinchmann}, {Charlot}, {White}, {Tremonti},
  {Kauffmann}, {Heckman} \& {Brinkmann}}{{Brinchmann}
  et~al.}{2004}]{Brinchmann2004}
{Brinchmann} J.,  {Charlot} S.,  {White} S.~D.~M.,  {Tremonti} C.,  {Kauffmann}
  G.,  {Heckman} T.,    {Brinkmann} J.,  2004, \mnras, 351, 1151

\bibitem[\protect\citeauthoryear{{Calzetti}, {Armus}, {Bohlin}, {Kinney},
  {Koornneef} \& {Storchi-Bergmann}}{{Calzetti} et~al.}{2000}]{Calzetti2000}
{Calzetti} D.,  {Armus} L.,  {Bohlin} R.~C.,  {Kinney} A.~L.,  {Koornneef} J.,
    {Storchi-Bergmann} T.,  2000, ApJ, 533, 682

\bibitem[\protect\citeauthoryear{{Cappellari}, {} \& {et al.}}{{Cappellari}
  et~al.}{2012}]{Cappellari2012}
{Cappellari} M.,  {}   {et al.} 2012, \nat, 484, 485

\bibitem[\protect\citeauthoryear{{Carpineti}, {Kaviraj}, {Darg}, {Lintott},
  {Schawinski} \& {Shabala}}{{Carpineti} et~al.}{2012}]{Carpineti2012}
{Carpineti} A.,  {Kaviraj} S.,  {Darg} D.,  {Lintott} C.,  {Schawinski} K.,
  {Shabala} S.,  2012, \mnras, p.~2262

\bibitem[\protect\citeauthoryear{{Cattaneo}, {} \& {et al.}}{{Cattaneo}
  et~al.}{2009}]{Cattaneo2009}
{Cattaneo} A.,  {}   {et al.} 2009, \nat, 460, 213

\bibitem[\protect\citeauthoryear{{Cen}}{{Cen}}{2012}]{Cen2012}
{Cen} R.,  2012, \apj, 755, 28

\bibitem[\protect\citeauthoryear{{Cole}, {Lacey}, {Baugh} \& {Frenk}}{{Cole}
  et~al.}{2000}]{Cole2000}
{Cole} S.,  {Lacey} C.~G.,  {Baugh} C.~M.,    {Frenk} C.~S.,  2000, MNRAS, 319,
  168

\bibitem[\protect\citeauthoryear{{Cox}, {Dutta}, {Di Matteo}, {Hernquist},
  {Hopkins}, {Robertson} \& {Springel}}{{Cox} et~al.}{2006}]{Cox2006}
{Cox} T.~J.,  {Dutta} S.~N.,  {Di Matteo} T.,  {Hernquist} L.,  {Hopkins}
  P.~F.,  {Robertson} B.,    {Springel} V.,  2006, \apj, 650, 791

\bibitem[\protect\citeauthoryear{{Croton}, {} \& {et al.}}{{Croton}
  et~al.}{2006}]{Croton2006}
{Croton} D.~J.,  {}   {et al.} 2006, \mnras, 365, 11

\bibitem[\protect\citeauthoryear{{Darg}, {} \& {et al.}}{{Darg}
  et~al.}{2010a}]{Darg2010a}
{Darg} D.~W.,  {}   {et al.} 2010a, MNRAS, 401, 1043

\bibitem[\protect\citeauthoryear{{Darg}, {} \& {et al.}}{{Darg}
  et~al.}{2010b}]{Darg2010b}
{Darg} D.~W.,  {}   {et al.} 2010b, \mnras, 401, 1552

\bibitem[\protect\citeauthoryear{{Davies}, {M{\"u}ller S{\'a}nchez}, {Genzel},
  {Tacconi}, {Hicks}, {Friedrich} \& {Sternberg}}{{Davies}
  et~al.}{2007}]{Davies2007}
{Davies} R.~I.,  {M{\"u}ller S{\'a}nchez} F.,  {Genzel} R.,  {Tacconi} L.~J.,
  {Hicks} E.~K.~S.,  {Friedrich} S.,    {Sternberg} A.,  2007, \apj, 671, 1388

\bibitem[\protect\citeauthoryear{{Deller} \& {Middelberg}}{{Deller} \&
  {Middelberg}}{2014}]{Deller2014}
{Deller} A.~T.,  {Middelberg} E.,  2014, \aj, 147, 14

\bibitem[\protect\citeauthoryear{{Ellison}, {Patton}, {Mendel} \&
  {Scudder}}{{Ellison} et~al.}{2011}]{Ellison2011}
{Ellison} S.~L.,  {Patton} D.~R.,  {Mendel} J.~T.,    {Scudder} J.~M.,  2011,
  \mnras, 418, 2043

\bibitem[\protect\citeauthoryear{{En{\ss}lin} \& {Gopal-Krishna}}{{En{\ss}lin}
  \& {Gopal-Krishna}}{2001}]{Ensslin2001}
{En{\ss}lin} T.~A.,  {Gopal-Krishna} 2001, \aap, 366, 26

\bibitem[\protect\citeauthoryear{{Fabian}}{{Fabian}}{1994}]{Fabian1994}
{Fabian} A.~C.,  1994, \araa, 32, 277

\bibitem[\protect\citeauthoryear{{Fabian}}{{Fabian}}{1999}]{Fabian1999}
{Fabian} A.~C.,  1999, \mnras, 308, L39

\bibitem[\protect\citeauthoryear{{Fabian}}{{Fabian}}{2012}]{Fabian2012}
{Fabian} A.~C.,  2012, \araa, 50, 455

\bibitem[\protect\citeauthoryear{{Fabian}, {Nulsen} \& {Canizares}}{{Fabian}
  et~al.}{1982}]{Fabian1982}
{Fabian} A.~C.,  {Nulsen} P.~E.~J.,    {Canizares} C.~R.,  1982, \mnras, 201,
  933

\bibitem[\protect\citeauthoryear{{Ferrarese} \& {Merritt}}{{Ferrarese} \&
  {Merritt}}{2000}]{Ferrarese2000}
{Ferrarese} L.,  {Merritt} D.,  2000, \apjl, 539, L9

\bibitem[\protect\citeauthoryear{{Gebhardt}, {} \& {et al.}}{{Gebhardt}
  et~al.}{2000}]{Gebhardt2000}
{Gebhardt} K.,  {}   {et al.} 2000, \apjl, 543, L5

\bibitem[\protect\citeauthoryear{{Granato}, {De Zotti}, {Silva}, {Bressan} \&
  {Danese}}{{Granato} et~al.}{2004}]{Granato2004}
{Granato} G.~L.,  {De Zotti} G.,  {Silva} L.,  {Bressan} A.,    {Danese} L.,
  2004, \apj, 600, 580

\bibitem[\protect\citeauthoryear{{G{\"u}ltekin}, {} \& {et al.}}{{G{\"u}ltekin}
  et~al.}{2009}]{Gultekin2009}
{G{\"u}ltekin} K.,  {}   {et al.} 2009, \apj, 698, 198

\bibitem[\protect\citeauthoryear{{Haehnelt}, {Natarajan} \& {Rees}}{{Haehnelt}
  et~al.}{1998}]{Haehnelt1998}
{Haehnelt} M.~G.,  {Natarajan} P.,    {Rees} M.~J.,  1998, \mnras, 300, 817

\bibitem[\protect\citeauthoryear{{Hardcastle}, {Evans} \&
  {Croston}}{{Hardcastle} et~al.}{2007}]{Hardcastle2007}
{Hardcastle} M.~J.,  {Evans} D.~A.,    {Croston} J.~H.,  2007, \mnras, 376,
  1849

\bibitem[\protect\citeauthoryear{{H{\"a}ring} \& {Rix}}{{H{\"a}ring} \&
  {Rix}}{2004}]{Haring2004}
{H{\"a}ring} N.,  {Rix} H.-W.,  2004, \apjl, 604, L89

\bibitem[\protect\citeauthoryear{{Hatton}, {Devriendt}, {Ninin}, {Bouchet},
  {Guiderdoni} \& {Vibert}}{{Hatton} et~al.}{2003}]{Hatton2003}
{Hatton} S.,  {Devriendt} J.~E.~G.,  {Ninin} S.,  {Bouchet} F.~R.,
  {Guiderdoni} B.,    {Vibert} D.,  2003, MNRAS, 343, 75

\bibitem[\protect\citeauthoryear{{Inskip}, {Tadhunter}, {Morganti}, {Holt},
  {Ramos Almeida} \& {Dicken}}{{Inskip} et~al.}{2010}]{Inskip2010}
{Inskip} K.~J.,  {Tadhunter} C.~N.,  {Morganti} R.,  {Holt} J.,  {Ramos
  Almeida} C.,    {Dicken} D.,  2010, \mnras, 407, 1739

\bibitem[\protect\citeauthoryear{{Iwasawa}, {Sanders}, {Evans}, {Trentham},
  {Miniutti} \& {Spoon}}{{Iwasawa} et~al.}{2005}]{Iwasawa2005}
{Iwasawa} K.,  {Sanders} D.~B.,  {Evans} A.~S.,  {Trentham} N.,  {Miniutti} G.,
     {Spoon} H.~W.~W.,  2005, \mnras, 357, 565

\bibitem[\protect\citeauthoryear{{Jahnke} \& {Macci{\`o}}}{{Jahnke} \&
  {Macci{\`o}}}{2011}]{Jahnke2011}
{Jahnke} K.,  {Macci{\`o}} A.~V.,  2011, \apj, 734, 92

\bibitem[\protect\citeauthoryear{{Jeong}, {Bureau}, {Yi}, {Krajnovi{\'c}} \&
  {Davies}}{{Jeong} et~al.}{2007}]{Jeong2007}
{Jeong} H.,  {Bureau} M.,  {Yi} S.~K.,  {Krajnovi{\'c}} D.,    {Davies} R.~L.,
  2007, \mnras, 376, 1021

\bibitem[\protect\citeauthoryear{{Kauffmann}, {} \& {et al.}}{{Kauffmann}
  et~al.}{2003}]{Kauffmann2003}
{Kauffmann} G.,  {}   {et al.} 2003, \mnras, 341, 33

\bibitem[\protect\citeauthoryear{{Kaviraj}, {} \& {et al.}}{{Kaviraj}
  et~al.}{007b}]{Kaviraj2007a}
{Kaviraj} S.,  {}   {et al.} 2007b, \apjs, 173, 619

\bibitem[\protect\citeauthoryear{{Kaviraj}, {} \& {et al.}}{{Kaviraj}
  et~al.}{2008}]{Kaviraj2008b}
{Kaviraj} S.,  {}   {et al.} 2008, MNRAS, 388, 67

\bibitem[\protect\citeauthoryear{{Kaviraj}, {} \& {et al.}}{{Kaviraj}
  et~al.}{2013}]{Kaviraj2013a}
{Kaviraj} S.,  {}   {et al.} 2013, \mnras, 429, L40

\bibitem[\protect\citeauthoryear{{Kaviraj}}{{Kaviraj}}{2009}]{Kaviraj2009a}
{Kaviraj} S.,  2009, \mnras, 394, 1167

\bibitem[\protect\citeauthoryear{{Kaviraj}}{{Kaviraj}}{2010}]{Kaviraj2010}
{Kaviraj} S.,  2010, \mnras, 406, 382

\bibitem[\protect\citeauthoryear{{Kaviraj}}{{Kaviraj}}{014a}]{Kaviraj2014a}
{Kaviraj} S.,  2014a, \mnras, 437, L41

\bibitem[\protect\citeauthoryear{{Kaviraj}}{{Kaviraj}}{014b}]{Kaviraj2014b}
{Kaviraj} S.,  2014b, \mnras, 440, 2944

\bibitem[\protect\citeauthoryear{{Kaviraj}, {Khochfar}, {Schawinski} \& {et
  al.}}{{Kaviraj} et~al.}{2008}]{Kaviraj2008}
{Kaviraj} S.,  {Khochfar} S.,  {Schawinski} K.,    {et al.} 2008, \mnras, 388,
  67

\bibitem[\protect\citeauthoryear{{Kaviraj}, {Kirkby}, {Silk} \&
  {Sarzi}}{{Kaviraj} et~al.}{007a}]{Kaviraj2007c}
{Kaviraj} S.,  {Kirkby} L.~A.,  {Silk} J.,    {Sarzi} M.,  2007a, \mnras, 382,
  960

\bibitem[\protect\citeauthoryear{{Kaviraj}, {Schawinski}, {Silk} \&
  {Shabala}}{{Kaviraj} et~al.}{011a}]{Kaviraj2011c}
{Kaviraj} S.,  {Schawinski} K.,  {Silk} J.,    {Shabala} S.~S.,  2011a, \mnras,
  415, 3798

\bibitem[\protect\citeauthoryear{{Kaviraj}, {Sohn}, {O'Connell}, {Yoon}, {Lee}
  \& {Yi}}{{Kaviraj} et~al.}{007c}]{Kaviraj2007b}
{Kaviraj} S.,  {Sohn} S.~T.,  {O'Connell} R.~W.,  {Yoon} S.,  {Lee} Y.~W.,
  {Yi} S.~K.,  2007c, \mnras, 377, 987

\bibitem[\protect\citeauthoryear{{Kaviraj}, {Tan}, {Ellis} \& {Silk}}{{Kaviraj}
  et~al.}{011b}]{Kaviraj2011}
{Kaviraj} S.,  {Tan} K.-M.,  {Ellis} R.~S.,    {Silk} J.,  2011b, \mnras, 411,
  2148

\bibitem[\protect\citeauthoryear{{Kewley}, {Groves}, {Kauffmann} \&
  {Heckman}}{{Kewley} et~al.}{2006}]{Kewley2006}
{Kewley} L.~J.,  {Groves} B.,  {Kauffmann} G.,    {Heckman} T.,  2006, \mnras,
  372, 961

\bibitem[\protect\citeauthoryear{{King}}{{King}}{2003}]{King2003}
{King} A.,  2003, \apjl, 596, L27

\bibitem[\protect\citeauthoryear{{Komatsu}, {} \& {et al.}}{{Komatsu}
  et~al.}{2011}]{Komatsu2011}
{Komatsu} E.,  {}   {et al.} 2011, \apjs, 192, 18

\bibitem[\protect\citeauthoryear{{Kondratko}, {} \& {et al.}}{{Kondratko}
  et~al.}{2006}]{Kondratko2006}
{Kondratko} P.~T.,  {}   {et al.} 2006, \apj, 638, 100

\bibitem[\protect\citeauthoryear{{Koss}, {Mushotzky}, {Veilleux} \&
  {Winter}}{{Koss} et~al.}{2010}]{Koss2010}
{Koss} M.,  {Mushotzky} R.,  {Veilleux} S.,    {Winter} L.,  2010, \apjl, 716,
  L125

\bibitem[\protect\citeauthoryear{{Ledlow}, {Owen}, {Yun} \& {Hill}}{{Ledlow}
  et~al.}{2001}]{Ledlow2001}
{Ledlow} M.~J.,  {Owen} F.~N.,  {Yun} M.~S.,    {Hill} J.~M.,  2001, \apj, 552,
  120

\bibitem[\protect\citeauthoryear{{Lintott}, {} \& {et al.}}{{Lintott}
  et~al.}{2008}]{Lintott2008}
{Lintott} C.~J.,  {}   {et al.} 2008, \mnras, 389, 1179

\bibitem[\protect\citeauthoryear{{Lonsdale}, {Diamond}, {Thrall}, {Smith} \&
  {Lonsdale}}{{Lonsdale} et~al.}{2006}]{Lonsdale2006}
{Lonsdale} C.~J.,  {Diamond} P.~J.,  {Thrall} H.,  {Smith} H.~E.,    {Lonsdale}
  C.~J.,  2006, \apj, 647, 185

\bibitem[\protect\citeauthoryear{{Lonsdale}, {Lonsdale}, {Smith} \&
  {Diamond}}{{Lonsdale} et~al.}{2003}]{Lonsdale2003}
{Lonsdale} C.~J.,  {Lonsdale} C.~J.,  {Smith} H.~E.,    {Diamond} P.~J.,  2003,
  \apj, 592, 804

\bibitem[\protect\citeauthoryear{{Madau} \& {Dickinson}}{{Madau} \&
  {Dickinson}}{2014}]{Madau2014}
{Madau} P.,  {Dickinson} M.,  2014, ArXiv e-prints

\bibitem[\protect\citeauthoryear{{Magorrian}, {} \& {et al.}}{{Magorrian}
  et~al.}{1998}]{Magorrian1998}
{Magorrian} J.,  {}   {et al.} 1998, \aj, 115, 2285

\bibitem[\protect\citeauthoryear{{Martin}, {} \& {et al.}}{{Martin}
  et~al.}{2005}]{Martin2005}
{Martin} D.~C.,  {}   {et al.} 2005, ApJ, 619, L1

\bibitem[\protect\citeauthoryear{{McConnell}, {} \& {et al.}}{{McConnell}
  et~al.}{2011}]{McConnell2011}
{McConnell} N.~J.,  {}   {et al.} 2011, \nat, 480, 215

\bibitem[\protect\citeauthoryear{{McNamara} \& {Nulsen}}{{McNamara} \&
  {Nulsen}}{2007}]{McNamara2007}
{McNamara} B.~R.,  {Nulsen} P.~E.~J.,  2007, \araa, 45, 117

\bibitem[\protect\citeauthoryear{{Morganti}, {Fogasy}, {Paragi}, {Oosterloo} \&
  {Orienti}}{{Morganti} et~al.}{2013}]{Morganti2013}
{Morganti} R.,  {Fogasy} J.,  {Paragi} Z.,  {Oosterloo} T.,    {Orienti} M.,
  2013, Science, 341, 1082

\bibitem[\protect\citeauthoryear{{Morrissey}, {} \& {et al.}}{{Morrissey}
  et~al.}{2007}]{Morrissey2007}
{Morrissey} P.,  {}   {et al.} 2007, \apjs, 173, 682

\bibitem[\protect\citeauthoryear{{Nenkova}, {Sirocky}, {Nikutta}, {Ivezi{\'c}}
  \& {Elitzur}}{{Nenkova} et~al.}{2008}]{Nenkova2008}
{Nenkova} M.,  {Sirocky} M.~M.,  {Nikutta} R.,  {Ivezi{\'c}} {\v Z}.,
  {Elitzur} M.,  2008, \apj, 685, 160

\bibitem[\protect\citeauthoryear{{Nesvadba}, {Lehnert}, {De Breuck}, {Gilbert}
  \& {van Breugel}}{{Nesvadba} et~al.}{2008}]{Nesvadba2008}
{Nesvadba} N.~P.~H.,  {Lehnert} M.~D.,  {De Breuck} C.,  {Gilbert} A.~M.,
  {van Breugel} W.,  2008, \aap, 491, 407

\bibitem[\protect\citeauthoryear{{Nesvadba}, {Polletta}, {Lehnert}, {Bergeron},
  {De Breuck}, {Lagache} \& {Omont}}{{Nesvadba} et~al.}{2011}]{Nesvadba2011}
{Nesvadba} N.~P.~H.,  {Polletta} M.,  {Lehnert} M.~D.,  {Bergeron} J.,  {De
  Breuck} C.,  {Lagache} G.,    {Omont} A.,  2011, \mnras, 415, 2359

\bibitem[\protect\citeauthoryear{{Nyland}, {} \& {et al.}}{{Nyland}
  et~al.}{2013}]{Nyland2013}
{Nyland} K.,  {}   {et al.} 2013, \apj, 779, 173

\bibitem[\protect\citeauthoryear{{Oke} \& {Gunn}}{{Oke} \&
  {Gunn}}{1983}]{Oke1983}
{Oke} J.~B.,  {Gunn} J.~E.,  1983, \apj, 266, 713

\bibitem[\protect\citeauthoryear{{Patton}, {Pritchet}, {Carlberg} \& {et
  al.}}{{Patton} et~al.}{2002}]{Patton2002}
{Patton} D.~R.,  {Pritchet} C.~J.,  {Carlberg} R.~G.,    {et al.} 2002, ApJ,
  565, 208

\bibitem[\protect\citeauthoryear{{Peng}}{{Peng}}{2007}]{Peng2007}
{Peng} C.~Y.,  2007, \apj, 671, 1098

\bibitem[\protect\citeauthoryear{{Robertson}, {Bullock}, {Cox}, {Di Matteo},
  {Hernquist}, {Springel} \& {Yoshida}}{{Robertson}
  et~al.}{2006}]{Robertson2006}
{Robertson} B.,  {Bullock} J.~S.,  {Cox} T.~J.,  {Di Matteo} T.,  {Hernquist}
  L.,  {Springel} V.,    {Yoshida} N.,  2006, \apj, 645, 986

\bibitem[\protect\citeauthoryear{{Rupke} \& {Veilleux}}{{Rupke} \&
  {Veilleux}}{2011}]{Rupke2011}
{Rupke} D.~S.~N.,  {Veilleux} S.,  2011, \apjl, 729, L27

\bibitem[\protect\citeauthoryear{{Schawinski}, {Dowlin}, {Thomas}, {Urry} \&
  {Edmondson}}{{Schawinski} et~al.}{2010}]{Schawinski2010}
{Schawinski} K.,  {Dowlin} N.,  {Thomas} D.,  {Urry} C.~M.,    {Edmondson} E.,
  2010, \apjl, 714, L108

\bibitem[\protect\citeauthoryear{{Schawinski}, {Thomas}, {Sarzi}, {Maraston},
  {Kaviraj}, {Joo}, {Yi} \& {Silk}}{{Schawinski} et~al.}{2007}]{Schawinski2007}
{Schawinski} K.,  {Thomas} D.,  {Sarzi} M.,  {Maraston} C.,  {Kaviraj} S.,
  {Joo} S.-J.,  {Yi} S.~K.,    {Silk} J.,  2007, \mnras, 382, 1415

\bibitem[\protect\citeauthoryear{{Schlegel}, {Finkbeiner} \&
  {Davis}}{{Schlegel} et~al.}{1998}]{Schlegel1998}
{Schlegel} D.~J.,  {Finkbeiner} D.~P.,    {Davis} M.,  1998, \apj, 500, 525

\bibitem[\protect\citeauthoryear{{Scott} \& {Kaviraj}}{{Scott} \&
  {Kaviraj}}{2014}]{Scott2014}
{Scott} C.,  {Kaviraj} S.,  2014, \mnras, 437, 2137

\bibitem[\protect\citeauthoryear{{Scudder}, {Ellison}, {Torrey}, {Patton} \&
  {Mendel}}{{Scudder} et~al.}{2012}]{Scudder2012}
{Scudder} J.~M.,  {Ellison} S.~L.,  {Torrey} P.,  {Patton} D.~R.,    {Mendel}
  J.~T.,  2012, \mnras, 426, 549

\bibitem[\protect\citeauthoryear{{Shabala} \& {Alexander}}{{Shabala} \&
  {Alexander}}{2009}]{Shabala2009}
{Shabala} S.,  {Alexander} P.,  2009, \apj, 699, 525

\bibitem[\protect\citeauthoryear{{Shabala}, {} \& {et al.}}{{Shabala}
  et~al.}{2012}]{Shabala2012}
{Shabala} S.~S.,  {}   {et al.} 2012, \mnras, 423, 59

\bibitem[\protect\citeauthoryear{{Shabala}, {Ash}, {Alexander} \&
  {Riley}}{{Shabala} et~al.}{2008}]{Shabala2008}
{Shabala} S.~S.,  {Ash} S.,  {Alexander} P.,    {Riley} J.~M.,  2008, \mnras,
  388, 625

\bibitem[\protect\citeauthoryear{{Silk} \& {Rees}}{{Silk} \&
  {Rees}}{1998}]{Silk1998}
{Silk} J.,  {Rees} M.~J.,  1998, \aap, 331, L1

\bibitem[\protect\citeauthoryear{Sivia}{Sivia}{1996}]{Sivia1996}
Sivia D.~S.,  1996, Data Analysis, A Bayesian Tutorial.
Oxford

\bibitem[\protect\citeauthoryear{{Somerville}, {Gilmore}, {Primack} \&
  {Dom{\'{\i}}nguez}}{{Somerville} et~al.}{2012}]{Somerville2012}
{Somerville} R.~S.,  {Gilmore} R.~C.,  {Primack} J.~R.,    {Dom{\'{\i}}nguez}
  A.,  2012, \mnras, 423, 1992

\bibitem[\protect\citeauthoryear{{Springel}, {Di Matteo} \&
  {Hernquist}}{{Springel} et~al.}{2005a}]{Springel2005a}
{Springel} V.,  {Di Matteo} T.,    {Hernquist} L.,  2005a, MNRAS, 361, 776

\bibitem[\protect\citeauthoryear{{Springel}, {Di Matteo} \&
  {Hernquist}}{{Springel} et~al.}{2005b}]{Springel2005}
{Springel} V.,  {Di Matteo} T.,    {Hernquist} L.,  2005b, \apjl, 620, L79

\bibitem[\protect\citeauthoryear{{Strateva}, {} \& {et al.}}{{Strateva}
  et~al.}{2001}]{Strateva2001}
{Strateva} I.,  {}   {et al.} 2001, \aj, 122, 1861

\bibitem[\protect\citeauthoryear{{Sturm}, {} \& {et al.}}{{Sturm}
  et~al.}{2011}]{Sturm2011}
{Sturm} E.,  {}   {et al.} 2011, \apjl, 733, L16

\bibitem[\protect\citeauthoryear{{Tabor} \& {Binney}}{{Tabor} \&
  {Binney}}{1993}]{Tabor1993}
{Tabor} G.,  {Binney} J.,  1993, \mnras, 263, 323

\bibitem[\protect\citeauthoryear{{Thomas}, {Maraston}, {Bender} \& {Mendes de
  Oliveira}}{{Thomas} et~al.}{2005}]{Thomas2005}
{Thomas} D.,  {Maraston} C.,  {Bender} R.,    {Mendes de Oliveira} C.,  2005,
  \apj, 621, 673

\bibitem[\protect\citeauthoryear{{Trager}, {Faber}, {Worthey} \& {Gonz{\'
  a}lez}}{{Trager} et~al.}{2000}]{Trager2000a}
{Trager} S.~C.,  {Faber} S.~M.,  {Worthey} G.,    {Gonz{\' a}lez} J.~J.,  2000,
  AJ, 119, 1645

\bibitem[\protect\citeauthoryear{{Tremonti}, {Heckman}, {Kauffmann},
  {Brinchmann}, {Charlot}, {White}, {Seibert}, {Peng}, {Schlegel}, {Uomoto},
  {Fukugita} \& {Brinkmann}}{{Tremonti} et~al.}{2004}]{Tremonti2004}
{Tremonti} C.~A.,  {Heckman} T.~M.,  {Kauffmann} G.,  {Brinchmann} J.,
  {Charlot} S.,  {White} S.~D.~M.,  {Seibert} M.,  {Peng} E.~W.,  {Schlegel}
  D.~J.,  {Uomoto} A.,  {Fukugita} M.,    {Brinkmann} J.,  2004, ApJ, 613, 898

\bibitem[\protect\citeauthoryear{{Veilleux} \& {Osterbrock}}{{Veilleux} \&
  {Osterbrock}}{1987}]{Veilleux1987}
{Veilleux} S.,  {Osterbrock} D.~E.,  1987, \apjs, 63, 295

\bibitem[\protect\citeauthoryear{{Wild}, {Heckman} \& {Charlot}}{{Wild}
  et~al.}{2010}]{Wild2010}
{Wild} V.,  {Heckman} T.,    {Charlot} S.,  2010, \mnras, 405, 933

\bibitem[\protect\citeauthoryear{{Wyder}, {} \& {et al.}}{{Wyder}
  et~al.}{2007}]{Wyder2007}
{Wyder} T.~K.,  {}   {et al.} 2007, \apjs, 173, 293

\bibitem[\protect\citeauthoryear{{Yang}, {Mo}, {van den Bosch}, {Pasquali},
  {Li} \& {Barden}}{{Yang} et~al.}{2007}]{Yang2007}
{Yang} X.,  {Mo} H.~J.,  {van den Bosch} F.~C.,  {Pasquali} A.,  {Li} C.,
  {Barden} M.,  2007, ApJ, 671, 153

\bibitem[\protect\citeauthoryear{{Yesuf}, {Faber}, {Trump}, {Koo}, {Fang},
  {Liu}, {Wild} \& {Hayward}}{{Yesuf} et~al.}{2014}]{Yesuf2014}
{Yesuf} H.~M.,  {Faber} S.~M.,  {Trump} J.~R.,  {Koo} D.~C.,  {Fang} J.~J.,
  {Liu} F.~S.,  {Wild} V.,    {Hayward} C.~C.,  2014, ArXiv e-prints

\bibitem[\protect\citeauthoryear{{Yi}}{{Yi}}{2003}]{Yi2003}
{Yi} S.~K.,  2003, ApJ, 582, 202

\bibitem[\protect\citeauthoryear{{Yi}, {Yoon}, {Kaviraj} \& {et al.}}{{Yi}
  et~al.}{2005}]{Yi2005}
{Yi} S.~K.,  {Yoon} S.-J.,  {Kaviraj} S.,    {et al.} 2005, ApJ, 619, L111

\bibitem[\protect\citeauthoryear{{Young}}{{Young}}{2002}]{Young2002}
{Young} L.~M.,  2002, \aj, 124, 788

\end{thebibliography}


\end{document}